\documentclass[fleqn, usenatbib]{mnras}

\usepackage{amsmath,amssymb, graphicx}
\usepackage{epstopdf}

\usepackage[caption=false]{subfig}
\usepackage{enumitem}

\usepackage{natbib}
\usepackage{comment}
\usepackage{color}

\date{Accepted XXX. Received YYY; in original form ZZZ}

\pubyear{2020}

\title[Mildly relativistic shocks -- part I]{MNRAS \LaTeXe\ Mildly relativistic magnetized shocks in electron-ion plasmas I. Electromagnetic shock structure}

\newcommand{\jn}{\textcolor{black}}
\newcommand{\jnn}{\textcolor{black}}
\newcommand{\jnr}{\textcolor{black}}

\newcommand{\al}{\textcolor{black}}

\newcommand{\mpo}{\textcolor{black}}
\newcommand{\Lsi}{\lambda_{\mathrm{si}}}
\newcommand{\Lse}{\lambda_{\mathrm{se}}}
\newcommand{\Ex}{E_{\mathrm{x}}}
\newcommand{\Ey}{E_{\mathrm{y}}}
\newcommand{\Ez}{E_{\mathrm{z}}}
\newcommand{\Bx}{B_{\mathrm{x}}}
\newcommand{\Bz}{B_{\mathrm{z}}}
\newcommand{\By}{B_{\mathrm{y}}}
\newcommand{\Bo}{B_{\mathrm{0}}}

\newcommand{\Wci}{\Omega_{\mathrm{ci}}}
\newcommand{\Wce}{\Omega_{\mathrm{ce}}}
\newcommand{\wpe}{\omega_{\mathrm{pe}}}
\newcommand{\degree}{^{o}}

\newcommand{\kbzx}{{k}_{\rm Bz, x}}

\newcommand{\kexx}{{k}_{\rm Ex, x}}
\newcommand{\kexy}{{k}_{\rm Ex, y}}
\newcommand{\gsh}{\gamma_{\rm sh}}

\title{Mildly relativistic magnetized shocks in electron-ion plasmas I. Electromagnetic shock structure}

\author[Ligorini et al.]{Arianna Ligorini,$^{1}$%\thanks{E-mail: arianna.ligorini@ifj.edu.pl}
Jacek Niemiec,$^{1}$\thanks{E-mail: jacek.niemiec@ifj.edu.pl}
Oleh Kobzar,$^{2}$
Masanori Iwamoto,$^{3}$
Artem Bohdan,$^{4}$
\newauthor 
Martin Pohl,$^{4,5}$
Yosuke Matsumoto,$^{6}$
Takanobu Amano,$^{7}$
Shuichi Matsukiyo,$^{3}$
Yodai Esaki,$^{8}$
\newauthor
and Masahiro Hoshino$^{7}$
\\
$^{1}$Institute of Nuclear Physics Polish Academy of Sciences, PL-31342 Krakow, Poland\\
$^{2}$Astronomical Observatory of the Jagiellonian University, PL-30244 Krakow, Poland\\
$^{3}$Faculty of Engineering Sciences, Kyushu University, Kasuga, Fukuoka, 816-8580, Japan\\
$^{4}$DESY, 15738 Zeuthen, Germany\\
$^{5}$Institute of Physics and Astronomy, University of Potsdam, 14476 Potsdam, Germany\\
$^{6}$Department of Physics, Chiba University, 1-33 Yayoi, Inage-ku, Chiba 263-8522, Japan\\
$^{7}$Department of Earth and Planetary Science, University of Tokyo, 7-3-1 Hongo, Bunkyo-ku, Tokyo 113-0033, Japan\\
$^{8}$Interdisciplinary Graduate School of Engineering Sciences, Kyushu University, Kasuga, Fukuoka, 816-8580, Japan
}

\begin{document}
\label{firstpage}
\pagerange{\pageref{firstpage}--\pageref{lastpage}}
\maketitle

\begin{abstract}
Mildly relativistic shocks in magnetized electron-ion plasmas are investigated with \mpo{2D kinetic particle-in-cell simulations of unprecedentedly high resolution and large scale for conditions \jnr{that may be found at} internal shocks in blazar cores.} Ion-scale effects cause corrugations along the shock surface \mpo{whose properties somewhat depend on the configuration of the mean perpendicular magnetic field, that is either} in or out of the simulation plane. We show that the synchrotron maser instability persists to operate in mildly relativistic shocks in agreement with theoretical predictions and produces coherent emission of upstream-propagating electromagnetic waves. Shock front ripples are excited in both mean-field configurations and they engender effective wave amplification. \jnr{The interaction of these waves with \mpo{upstream} plasma generates electrostatic wakefields.}
%that can energize electrons.
\end{abstract}

\begin{keywords}
acceleration of particles, instabilities, galaxies:jets, methods:numerical, plasmas, shock waves
\end{keywords}

\section{Introduction}
The origin of energetic particles is a long-standing problem of major importance in astrophysics. 
While it is widely assumed that cosmic rays (CRs) with energies up to $\sim10^{15}$~eV are produced at non-relativistic shocks of Galactic supernova remnants, higher-energy particles, in particular the so-called ultra-high-energy cosmic rays (UHECRs) with energies above $\sim10^{18}$ eV, are presumably generated in extragalactic systems with relativistic plasma outflows -- active galactic nuclei (AGN) and/or gamma-ray bursts (GRBs). Non-thermal synchrotron and inverse Compton emission in blazar jets extends in broad energy range from radio up to TeV $\gamma$ rays, indicating the presence of ultrarelativistic electrons. Recently established possible association of one of the high-energy neutrino sources with a flaring blazar TXS 0506+056 \citep{2018Sci...361.1378I} shows that also CR hadrons can be produced in AGN. 

High-energy particles in AGN and GRBs are often assumed to be accelerated at shock waves associated with the jets. These shocks have Lorentz factors, $\gamma_{\rm sh}$, ranging from mildly-relativistic to ultrarelativistic values. Many such systems are magnetized, exhibiting inherently quasi-perpendicular and superluminal conditions. 
%Such 
\jnr{Superluminal} shocks are mediated by magnetic reflection of the incoming flow off the shock-compressed magnetic field \cite[e.g.][]{langdon1988,gallant1992,hoshino1992}. Coherent gyration of particles at the shock front breaks up in bunches of charge and triggers the synchrotron maser instability (SMI), which excites large-amplitude electromagnetic waves of the extraordinary mode (X-mode) that can escape towards the upstream region. This precursor wave emission has been confirmed through one-dimensional (1D) 
\citep[e.g.][]{langdon1988,hoshino1991,gallant1992,hoshino1992,amato2006,plotnikov2019} and two-dimensional (2D) \citep[e.g.][]{sironi2009,sironi2011,iwamoto2017,iwamoto2018,plotnikov2018,iwamoto2019} PIC simulations. 
In the electron-ion plasmas, interactions of the incoming electrons with the precursor waves can also generate large-amplitude longitudinal electrostatic oscillations, so-called wakefield \citep{lyubarsky2006}. As demonstrated by \citet{hoshino2008}, a large-amplitude coherent electromagnetic wave propagating in the plasma can expel electrons in front of the wave packet and so induces a longitudinal polarization electric field. Electron expulsion \mpo{results because} the so-called ponderomotive force is proportional to the gradient of the wave pressure and acts much stronger on electrons than ions. The electric field excites longitudinal electron motions that lead to the electrostatic Langmuir waves.
The formation of large-amplitude wakefields results from the parametric decay instability \citep[PDI; e.g.][]{kruer1988}. In this wave-wave interaction the large-amplitude electromagnetic (pump) wave decays into a Langmuir wave and a scattered electromagnetic (light) wave.
If the pump-wave frequency is much larger than the plasma frequency, Forward Raman Scattering (FRS) is triggered, in which the scattered electromagnetic wave and the Langmuir wave propagate in the same direction as the pump wave. The wavelength of the Langmuir wave is close to the electron inertial length, and its phase velocity approaches the group velocity of the pump wave, that is close to the speed of light.
Electrons and ions can be energized to very high energies in a manner analogous to wakefield acceleration (WFA) during the nonlinear \mpo{collapse} of the Langmuir waves \citep{hoshino2008}. 
WFA was first proposed in laboratory plasmas \citep{1979PhRvL..43..267T} and later applied to UHECR acceleration \citep[e.g.][]{2002PhRvL..89p1101C}. It was then demonstrated through laser plasma experiments and simulations \citep[e.g.][]{kuramitsu2008} that the WFA produces power-law energy spectra with a spectral index of~2.

Relativistic magnetized shocks have recently been studied with 2D PIC simulations for the case of pair plasmas \citep{iwamoto2017,iwamoto2018,sironi2009,plotnikov2018}, electron-ion plasmas \citep{sironi2011,stockem2012,iwamoto2019} and also mixed-composition plasmas \citep{stockem2012}. 
\citet{iwamoto2017} \mpo{demonstrated} that simulations need to have high numerical resolution to capture the precursor waves, in which case coherent waves persist even in weakly magnetized plasmas, dominated by the relativistic Weibel instability \cite[e.g.][]{Kato2010,sironi2011}. In pair plasmas, the precursor wave amplitudes were found to be systematically smaller in 2D simulations \mpo{than in the 1D case, but are still} sufficient to disturb the upstream medium. 
2D simulations with magnetic field in the simulation plane showed that also ordinary mode (O-mode) waves are excited, \mpo{which at low magnetizations are amplified by} the Weibel instability \citep{iwamoto2018}. 
The amplitudes in pair plasmas are in general much smaller than at ion-electron shocks \citep{iwamoto2019}. In conditions of high electron magnetization the wave energy exceeds that in pair plasmas by almost two orders of magnitude, and the 2D amplitude is close to the 1D level. This amplification at high-\jn{$\gsh$} shocks is attributed to a positive feedback process associated with the ion-electron coupling through the induced wakefields. In the turbulent wakefields close to the shock the electrons can be efficiently heated so that the energy equipartition between electrons and ions may be achieved before the flow arrives at the shock front. At the same time non-thermal electrons and ions can be generated. 

Most \mpo{published studies address} ultra-relativistic shocks with Lorentz factors $\gsh \geq 10$. The mildly relativistic regime, $\gsh \approx 3$, has \mpo{been explored only with low-resolution studies which \jnr{for superluminal shocks} show very weak \citep{sironi2011} or no wakefield \citep{lyubarsky2006}.} It has been estimated that only for electron-ion shocks with $\gsh\gtrsim (m_i/m_e)^{1/3}$, where $m_i/m_e$ is the ion-to-electron mass ratio, the electrons will form ring-like phase-space distribution unstable to SMI. \mpo{If so, one would expect little electron energization upstream of the shock in blazar jets, which has important consequences for their synchrotron and inverse Compton emission} \cite[e.g.][]{sikora2013}.

Here we \mpo{revisit the efficiency of WFA and the level of the electron-proton coupling at mildly-relativistic 
\jnr{magnetized}
%strictly perpendicular 
shocks in electron-ion plasma with unprecedentedly high-resolution 2D PIC simulations}. We also account for ion-scale corrugations of the shock surface by employing a very large computational box.
\jnr{We study strictly perpendicular shocks, in which the strength of the precursor wave is expected to be largest for all superluminal obliquities \citep{lyubarsky2006,sironi2011}.}
%We focus on conditions \mpo{typical for internal shocks, namely a shock Lorentz factor of $\gamma_{sh}\simeq 3$ and the} plasma magnetization (the ratio of the Poynting flux to the kinetic energy flux) $\sigma=0.1$.
\jnr{We assume} a shock Lorentz factor of $\gamma_{sh}\simeq 3$ and the plasma magnetization (the ratio of the Poynting flux to the kinetic energy flux) $\sigma=0.1$. \jnr{These values are in the range of those expected for internal shocks in AGN jets.}
In this first paper we discuss the shock structure and the generation of plasma instabilities and waves. In a forthcoming publication (Ligorini et al., in preparation, Paper II) we present the particle acceleration and heating mechanisms and discuss the energy transfer from ions to electrons downstream of the shock. Section~\ref{sec:setup} presents the simulation setup. 
Section~\ref{sec:out-of-plane} shows results for the out-of-plane field orientation, which are compared to the case with the in-plane magnetic field in Section~\ref{sec:in-plane}. Section~\ref{sec:summary} presents a summary and conclusions of this first part of our study.

\section{Simulation setup} 
\label{sec:setup}
\begin{figure}
\includegraphics[width=0.99\linewidth]{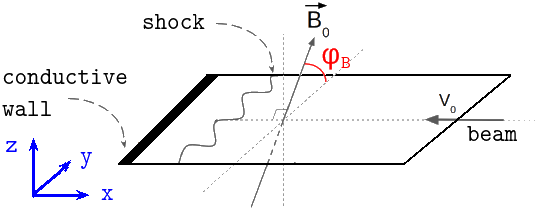}
\caption{Illustration of the simulation setup.}
\label{fig:setup}
\end{figure}

We use a modified version of the relativistic electromagnetic PIC code TRISTAN \citep{buneman1993} with MPI-based parallelization \citep{niemiec2008} and the option to trace individual particles. The simulation setup is shown in Fig.~\ref{fig:setup}. \mpo{An electron-ion beam flows with speed $\boldsymbol{v_0}$
in the negative $x$-direction. It bounces off a {\itshape reflective wall} at the left side of the simulation box and collides} with the incoming flow to form a shock propagating in the positive $x$-direction.

\mpo{In our 2D3V simulations we use a two-dimensional spatial grid but follow three components of particle momenta and electromagnetic fields.}
The beam carries a large-scale homogeneous magnetic field, $\boldsymbol{B_0}$, \mpo{oriented perpendicular to the shock normal, and the associated motional electric field, $\boldsymbol{E_0} = - \boldsymbol{\mathit v_0 \times B_{0}}$.} We study two configurations of the large-scale field with respect to the simulation plane, \jn{described by the angle~$\varphi_B$ (see Fig.~\ref{fig:setup})}: the \mpo{
{\it out-of-plane} magnetic field, $\boldsymbol{B_0}=\jn{B_{0z}}\boldsymbol{\hat{z}}$ and \jn{$\varphi_B=90\degree$} and the {\it in-plane} setup with $\boldsymbol{B_0}=\jn{B_{0y}}\boldsymbol{\hat{y}}$ and \jn{$\varphi_B=0\degree$}.}

\mpo{The beam Lorentz factor, \jn{$\gamma_{0}=2.03$}, results in the shock Lorentz factor $\gamma_{\rm sh}\simeq 3.3$ in the upstream rest frame. The total plasma magnetization, $\sigma = 0.1$, is written with simulation-frame magnetic-field strength, $B_{0}$, and ion density, $N_{i}$, as}
%as expected in AGN shocks, 
$\sigma = B_{0}^2/(\mu_{0} N_{i}(m_{e} +m_{i})\gamma_{0}c^2)$, where $c$ is the speed of light, $\mu_{0}$ is the permeability of free space, $m_e$ and $m_i$ are the electron and ion mass, respectively \citep{hoshino1992}. \mpo{The reduced ion-to-electron mass ratio, $m_{i}/m_{e} = 50$, determines the electron and ion magnetizations, $\sigma_{e}\simeq 5.1$ and $\sigma_{i}\simeq \sigma$, through $1/\sigma = 1/\sigma_{e} + 1/\sigma_{i}$.}
{We verified that our results do not change if a higher mass ratio of $m_{i}/m_{e}$ = 100 is used.}  

The unit of length used here is the ion skin depth, $\Lsi = c/\omega_{\rm pi}$, where $\omega_{\rm pi}=\sqrt{e^2N_i/\gamma_0\epsilon_0m_{i}}$ is the {relativistic} ion plasma frequency. {Here, $e$ is the electron charge, and $\epsilon_0$ is the vacuum permittivity. Time is expressed in units of the upstream ion cyclotron frequency $\Wci = (e\Bo)/(m_i \gamma_0)$. We ran our 2D simulations up to $t_{\rm max}=84.3\,\Wci^{-1}$ and complementary 1D simulations reach $t_{\rm max}=163.1\,\Wci^{-1}$. The time-step is $\delta t={1/1131}\,\omega_{\rm pi}^{-1}={1/3556.8}\Wci^{-1}$.} 

\mpo{\citet{iwamoto2017} noted that numerical investigations of magnetized shocks require high resolution, otherwise the precursor waves 
may be artificially damped. {Based on extensive tests described in {Appendix \ref{app:conv_test}, we }set the grid resolution to} 
%{$\Delta x = 1/80 \Lse = 1/566 \Lsi $}  
$\Lse = 80\Delta$, where $\Lse=\sqrt{m_e/m_i}\Lsi$ is the electron skin depth and $\Delta$ is the size of the grid cells. The corresponding ion skin depth is $\Lsi\simeq 566\Delta$. This resolution is twice larger than that adopted in \citet{iwamoto2017,iwamoto2018}.   
This unprecedentedly high resolution allowed us to detect precursor waves in the mildly-relativistic regime, that were invisible with} lower resolution \citep[e.g.][]{sironi2011}. Since our convergence tests show no dependence of the results on the number of particles per cell, $N_{\rm ppc}$, here we use $N_{\rm ppc}=10$ per particle species.

Relativistic shock simulations are extremely prone to the numerical Cherenkov instability that artificially heats and slows down the plasma beam \citep{yee1966,birdsall1991,hockney1981}. We minimize these unphysical effects, by using Friedman filters, a fourth-order accurate FTFD field-pusher \citep{greenwood2004}, and also by injecting cold plasma. 
\jnr{This numerical model also stabilizes the beam against the so-called finite-grid instability arising from an unresolved Debye length. The performance of the model has been extensively verified through test simulations.}
\mpo{Our {\itshape moving injection layer} makes sure that the simulated plasma contains all particles and waves propagating upstream, while we minimize the propagation time of the unperturbed beam.}

The transverse size of our simulation box is $L_y=5760\Delta\simeq 10\Lsi$,
\mpo{enough to capture structures in the shock surface with a characteristic length of several $\Lsi$. The box length, $L_x$, increases during simulations and reaches a final size of $L_x=160000\Delta\simeq 283\Lsi$.}

\section{\mpo{Shocks with out-of-plane magnetic field}}
\label{sec:out-of-plane}

\mpo{A mildly relativistic strictly perpendicular shock in ion-electron plasma {with large-scale} magnetic field pointing out of the simulation plane is followed up to $t_{\rm max}\Wci= 84.3$. Unlike for highly relativistic {flows} \citep[e.g.][]{sironi2011, iwamoto2019}, {the mildly relativistic shock is not} laminar. At $t\Wci\sim 6.5$ it develops corrugations visible in both the density and the {electromagnetic} field,
that are fully developed at $t\Wci\sim 8.5$.}
In Section~\ref{sec:linstage} we first present the structure of the semi-laminar shock at $t\Wci\simeq 7.5$ 
{to demonstrate that the SMI already operates at this early stage {in line with theoretical expectations}.
{In Section~\ref{sec:latestage}} \mpo{we discuss the} fully-evolved rippled shock.} 

\begin{figure}
%\begin{center}
\centering
\includegraphics [width=0.99\linewidth] {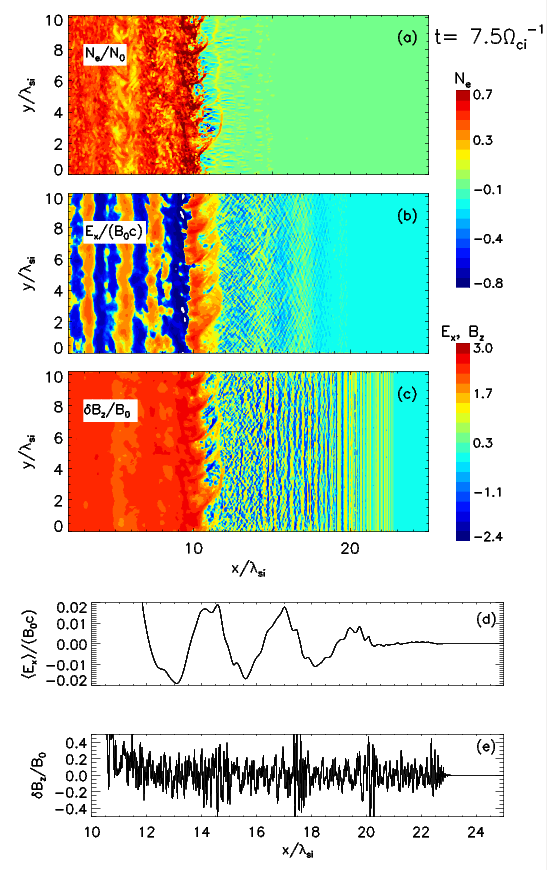}
%\includegraphics [width=0.99\linewidth] {plot/fig2.png}
%\end{center}
\caption{Distribution at time $t\Wci = 7.5$ of the normalized electron density, $N_{\mathrm{e}}/N_{0}$ (a), and \mpo{a component of the electric field, $E_{x}$ (b), and the magnetic field fluctuations, $\delta B_{z}$ (c). Logarithmic scaling is applied, which is sign-preserving  for electromagnetic fields (e.g. $\mathrm{sgn}(B_{z}) \cdot \{2+\log[\max(|B_{z}|/B_{0},10^{-2})]\}$),} so that field amplitudes below $10^{-2}B_{0}$ are not resolved. Panel d) shows the transversely averaged profile of the electric field, $\langle\Ex\rangle$, and panel e) displays the profile of $\delta\Bz$ taken along $y/\Lsi=6$.}
\label{fig:structure_early}
\end{figure}

\begin{figure}
\begin{center}
{\includegraphics[width=0.9\linewidth]{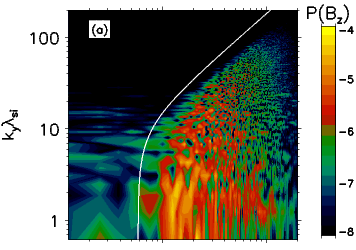}} 
{\includegraphics[width=0.9\linewidth]{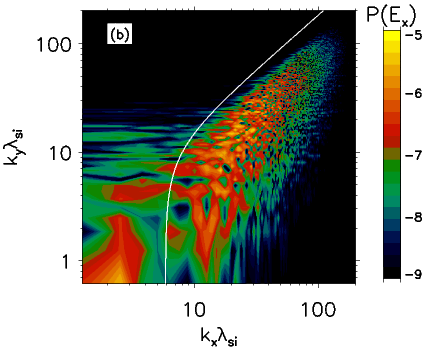}} 
\end{center}
\caption{Fourier power spectra for $\Bz$ \mpo{and $\Ex$ at $t = 7.5 \Wci^{-1}$, calculated upstream of the shock in the region $x/\Lsi=13-18$ (compare Fig.~\ref{fig:structure_early}). {The solid white line represents the theoretical cutoff of the precursor waves.}}
}
\label{fig:fourier_early}
\end{figure}

\begin{figure*}
\begin{center}
{\includegraphics[scale=0.47]{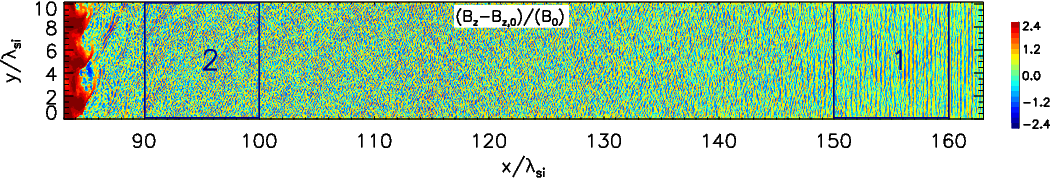}} 
%{\includegraphics[scale= 0.5]{plot/fig4a.png}} 
\end{center}
\caption{Map of normalized magnetic-field fluctuations, $\delta\Bz$, at time $t\Wci = 56.2$. The shock is located at $x/\Lsi\simeq 84$. Sign-preserving logarithmic scaling is applied (see Fig.~\ref{fig:structure_early}). Regions~1 and~2 highlighted with blue squares mark the initial positions of the regions chosen for the Fourier-Laplace spectra presented in Figs.~\ref{fig:omegak_bz_oop_1} and~\ref{fig:omegak_bz_oop_2}. \jnr{Note that we show only part of the precursor, that extends up to $x/\Lsi\approx 177\approx ct$.}}
\label{fig:full_map200}
\end{figure*}

\subsection{Semi-laminar shock stage}
\label{sec:linstage}

\mpo{Fig.~\ref{fig:structure_early} displays the initial {development of} the shock front, then located at $x\simeq 11\Lsi$. The density compression by a factor around $3$ is the theoretically expected value, $\kappa=3.18$, for relativistic plasma with adiabatic index $\Gamma={2}$ \citep{gallant1992}. 
Upstream of the shock, at $x\gtrsim 12\Lsi$, one can see X-mode waves as plane-wave fluctuations in $B_z$ that move with the speed of light and have a wave vector $\boldsymbol{k}_{\rm Bz}={k}_{\rm Bz, x}\boldsymbol{\hat{x}}$. 
\jnr{The tip of the waves is at $x\approx 23\Lsi$, which is the light travel distance $ct$ from the reflective wall for $t\Wci = 7.5$.}
One can also see co-moving longitudinal fluctuations in $\Ex$ at longer wavelength (Figs.~\ref{fig:structure_early}(b) and \ref{fig:structure_early}(d))
with the same phase velocity. The normalized amplitude of these {\it electrostatic} waves averaged over the three oscillations observed, $E_x/B_0c\simeq 1.8\cdot 10^{-2}$, is a factor ten smaller than that of the X-mode waves. Note, that already at this
very early stage the shock surface is perturbed.}

The emission of X-mode \mpo{waves indicates the operation of SMI at the shock \citep[e.g.][]{hoshino2008,iwamoto2017,iwamoto2018}. We calculated Fourier spectra of $\Bz$ and $\Ex$ for a region upstream of the shock at $x/\Lsi=13-18$ (see Fig.~\ref{fig:structure_early}). 
\jnr{The waves localized in the $x/\Lsi\sim 20-23$ region were emitted during the initial beam reflection off the conducting wall, when the shock had not yet formed. They are heavily affected by the initial conditions and hence not considered in our analysis.}
The X-mode waves can reach the precursor only, if the $x$-component of their group velocity is faster than that of the shock, which imposes a limit to the wave vector. Fig.~\ref{fig:fourier_early} demonstrates that most of the wave power is indeed observed at a $k_x$ larger than that limit, supporting the association of the X-mode waves with SMI at the shock. }

\mpo{The interaction of the precursor waves with the magnetized 
electron-proton plasma upstream should lead to electrostatic wakefield fluctuations that are evident at $k_{\rm Ex,x}\Lsi\simeq 2.5-3.0$ in the power spectrum of $\Ex$ in
Fig.~\ref{fig:fourier_early}(b).} 
The signal at \jn{$\kexx\Lsi\sim 1-4$ and $\kexy\Lsi\sim 3-20$} is due to filamentation and discussed in detail in Section~\ref{sec:filam}.

\mpo{In Appendix~\ref{app:lintheory} we derive the expected frequency of the X-mode waves as $\omega'/\Wce\gtrsim 1$, where $\Wce\simeq 2.25\wpe$ is the electron cyclotron frequency and the prime denotes a quantity measured in the upstream frame. Then, wakefield generation by Raman scattering should yield $\omega'_L\simeq \wpe$ and $k'_L \simeq 1/\Lse$ \citep{kruer1988, hoshino2008}. In the downstream (simulation) frame $k_{L,y}=k_{L,y}'$ and
\begin{equation}
\begin{aligned}
k_{L,x}=&\gamma_0~ k'_{L,x}~(1-\beta_0\frac{\omega'_L}{c~ k'_{L,x}}) \simeq \gamma_0 ~ \frac{1}{\Lse}(1-\beta_0 ~ \frac{\wpe~ \Lse}{c}) \\
=& \frac{\gamma_0}{\Lsi}~ \sqrt{\frac{m_i}{m_e}}~(1-\beta_0) \simeq \frac{1.9}{\Lsi},
\end{aligned}
\label{eq:exwavelenght}
\end{equation}
where we inserted our parameters to derive the last expression. Despite the poor wavenumber sampling of the signal in Fig.~\ref{fig:structure_early}, the match is reasonable. }

\subsection{Non-linear \jn{electromagnetic} shock structure}
\label{sec:latestage}
\subsubsection{Precursor waves}
\begin{figure}
\begin{center}
\includegraphics[width=0.99\linewidth]{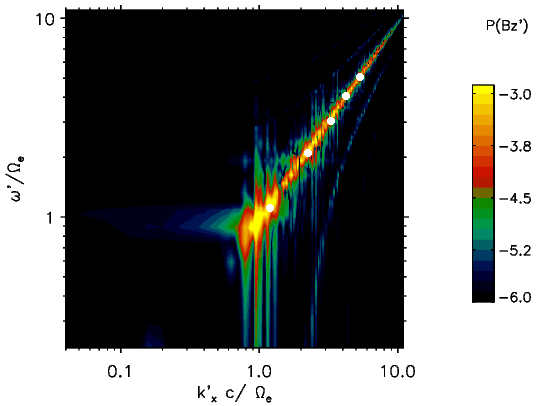}
\end{center}
\caption{\mpo{Fourier-Laplace spectrum in the \emph{upstream} frame of $B_z'$} in $\omega' - k'_x$ space taken along $y=5\Lsi$ in Region~1 marked in Fig.~\ref{fig:full_map200}. The time interval is $0.562\,\Wci^{-1}\simeq28.1\,\Wce^{-1}$, starting from time $t\Wci\simeq 56.23$. The angular frequency, $\omega'$, and the wave vector, $k'_x$, are normalized with the electron cyclotron frequency, $\Wce$. Overlaid with {white} dots is the dispersion relation derived in Appendix~\ref{app:lintheory}.
}
\label{fig:omegak_bz_oop_1}
\end{figure}

\begin{figure}
\begin{center}
\includegraphics[width=0.99\linewidth]{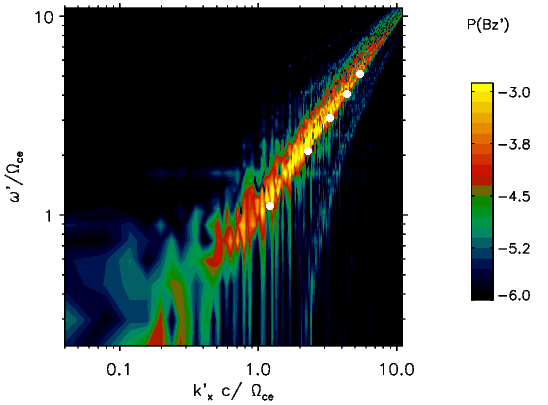}
\end{center}
\caption{\mpo{Fourier-Laplace spectrum of $B_z'$ at $y=5\Lsi$ as in Fig.~\ref{fig:omegak_bz_oop_1}, but here for Region~2 marked in Fig.~\ref{fig:full_map200}.}}
\label{fig:omegak_bz_oop_2}
\end{figure}

\begin{figure}
\begin{center}
\includegraphics [width=0.97\linewidth] {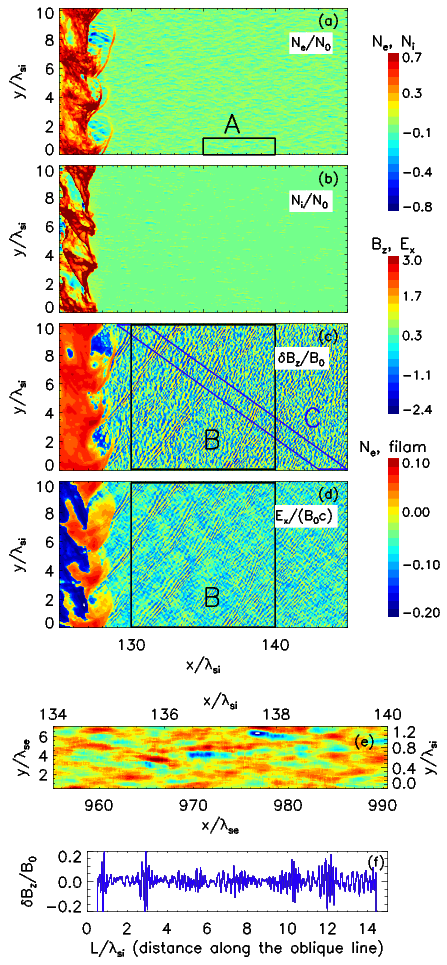}
\end{center}
\caption{Normalized \mpo{electron, (a), and ion density, (b), as well as the fluctuations in $\delta\Bz$ (c) and $\Ex$ (d) at $t\Wci = 84.3$. Logarithmic scaling is applied as in Fig.~\ref{fig:structure_early}. Panel (e) shows a close-up of the electron density in Region~A. Region~C in panel (c) is chosen to plot in panel (f) a profile of $\delta\Bz$.} }
\label{fig:full_map}
\end{figure}

Fig.~\ref{fig:full_map200} demonstrates that at time $t\Wci = 56.2$ the magnetic-field fluctuations extend to the far upstream. 
There, at $x/\Lsi \gtrsim 148$, one can only find waves 
\jnr{that have been emitted very early when the shock was semi-laminar shock and have a  very large $x$-component of their group speed.}
Hence the precursor waves retain their plane-wave character. \jnr{Behind this region,} closer to the shock, the waves also have an oblique component. Near the shock and up to $x/\Lsi\sim 120$ (see also Fig.~\ref{fig:full_map}) 
%the 
\jnr{these oblique}
waves form a quasi-regular pattern of oblique stripes.
\jnr{The emergence of the oblique wave component and formation of oblique stripes}
%that 
are related to the ripples in the shock surface, as we demonstrate below. A~Fourier-Laplace analysis in two selected regions of the shock precursor confirms that the waves \jnr{in the upstream region} are X modes, presumably generated through SMI. {The regions are stationary in the upstream plasma rest frame, and their initial location in the simulation box is marked in Fig.~\ref{fig:full_map200}}. The electromagnetic field is correspondingly transformed to the upstream frame. The Fourier-Laplace power spectra \jnn{are shown in Figs.~\ref{fig:omegak_bz_oop_1} and~\ref{fig:omegak_bz_oop_2}. They}
can be
compared with the theoretical dispersion relation \jn{for the electron SMI} discussed in Appendix~\ref{app:lintheory}.
\jnn{Since the group velocity of the waves emitted at the shock is small except for the wave numbers near the light mode, the majority of the waves cannot outrun the shock and propagate upstream, and hence the dispersion relation is indicated with white dots only along $\omega' = k' c$ in Figs.~\ref{fig:omegak_bz_oop_1} and~\ref{fig:omegak_bz_oop_2}.}

\mpo{The low-frequency modes of SMI generated by ions are mostly subluminous \jn{and do not propagate ahead of the shock}. In any case, they would not be detectable with our time window, $t\Wci\simeq 56.23$ to $t\Wci\simeq56.79$, and the sampling every 10 time steps.}
 
\mpo{Fig.~\ref{fig:omegak_bz_oop_1} demonstrates that in \emph{Region~1} far ahead of the shock the observed signal very well matches the theoretical dispersion relation for the electron SMI. In particular, the wave power is mostly along the light mode, $\omega'=k'c$, and a few harmonic modes exist for a wide wave vector range. The signal between the harmonics arises from fluctuations in the shock-compressed magnetic field.
The power spectrum in \emph{Region~2}, shown in Fig.~\ref{fig:omegak_bz_oop_2}, is heavily influenced by shock rippling and also by the non-linear evolution of the wave modes, but retains qualitative agreement with the electron maser model.} \jnr{Between Region 1 and 2 one finds a slow transition from parallel to oblique modes that represents a spatial mapping of the temporal development of shock rippling.}

\subsubsection{Effects of shock rippling on wave properties \label{sec:ripples}}
The shock ripples visible in Figs.~\ref{fig:full_map200} and~\ref{fig:full_map} \mpo{propagate in $-y$-direction with average speed
\jn{$v_{\rm rippl}\approx 0.8c$},
commensurate with that of ions gyrating at the shock. Their mean separation {along the shock surface}, $\lambda_{\rm rippl}\simeq 3.3\Lsi\simeq 2\,r_g$, and extension in $x$-direction, $1.6\Lsi\simeq r_g$, {reflect the} ion gyro-radius, $r_g$.} We associate the shock ripples with the modulation of shock-reflected ions along the shock surface first described by \citet{burgess2007} for low-Mach-number nonrelativistic shocks. The instability occurs only in simulations with out-of-plane magnetic field, \mpo{for which the ions gyrate in the simulation plane. In contrast, parallel-propagating waves driven by ion temperature anisotropy are frequently observed with in-plane field in the regime of low Mach numbers} \citep[e.g.][]{winske1988,2014PhPl...21b2102U}, but also 
in studies of high-Mach-number perpendicular nonrelativistic shocks \citep{wieland2016} and ultrarelativistic perpendicular shocks with moderate magnetizations, $\sigma\lesssim 0.1$ \citep{sironi2013}.  

In our simulation with $\varphi_B=90^{\circ}$ 
the shock ripples quickly grow from small-scale fluctuations to a long-wave mode visible in Figs.~\ref{fig:full_map200} and~\ref{fig:full_map}, \mpo{in particular in the ion density.} Their structure is highly dynamic on time-scales shorter than $\Wci^{-1}$ \mpo{and driven by the magnetic-field compression and charge separation induced by the} different inertia of electrons and ions. \mpo{Arcs of increased magnetic field, electron density, and associated electric field are generated. The maps of $B_z$ and $E_x$ in Figs.~\ref{fig:full_map}(c-d) suggest that
these arcs are the origin of} the observed pattern of oblique waves.

\begin{figure}
\includegraphics[width=0.99\linewidth]{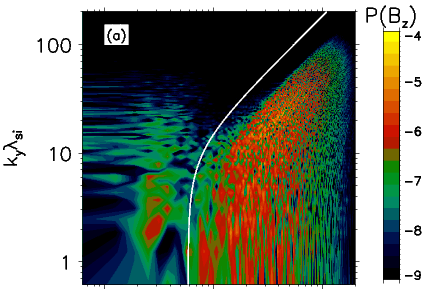}
\includegraphics[width=0.99\linewidth]{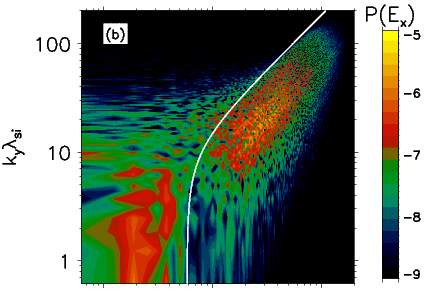}
\includegraphics[width=0.99\linewidth]{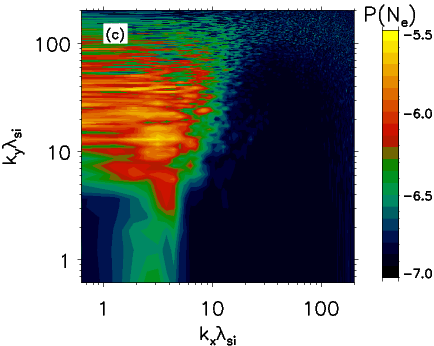} 
\caption{Fourier spectra for $\Bz$ fluctuations (a), $\Ex$ (b), and the electron density (c) for Region B at $x/\Lsi=130-140$ at time $t\Wci = 84.3$ (see Fig.~\ref{fig:full_map}). The solid white line represents the wave vector cutoff of precursor waves.}% \mpo{WHY DO WE HAVE $E_x \gg B_z$ NEAR THE WHITE LINE?? }}
\label{fig:fourier2}
\end{figure}

\mpo{The oblique structure of the precursor waves result from relativistic retardation and aberration of light and precursor-wave emission in a direction normal to the local front of the arcs. The arcs move with $v_{\rm rippl}$ close to $c$ in the negative $y$-direction. Retardation of emission in $x$-direction in the simulation frame gives rise to the stripes of high wave intensity seemingly oriented at an angle of $\vartheta\approx 37 ^{\circ}$ with the $y$-axis. Aberration provides a large $x$-component of the wave speed, so that the precursor waves can outpace the shock.}
The emission normal to the arc front results \mpo{from phase bunching \jn{of the electron distribution} \citep{hoshino1991,sprangle1977} which requires that the frequency of the wave be} slightly higher than the plasma cyclotron frequency. 
\jn{Then,} the particles on average gyrate less than $2\pi$ in a wave period and slip behind the waves. After a number of wave periods their distribution \jn{is} bunched \jn{in gyrophase. The wave emission thus will be defined by the structure of} the compressed magnetic field at the arcs, in which the electrons gyrate.
The \jn{combined effects of the gyrophase bunching, retardation, and aberration} cause the direction of \jn{precursor wave} emission to vary with the evolving shape of the ripples, leading to a wide range of angles, as visible in Figs.~\ref{fig:full_map200} and~\ref{fig:full_map}(c-d) and apparent in 2D Fourier power spectra of fluctuations in $\Bz$ and $\Ex$ shown in Figs.~\ref{fig:fourier2}(a-b) for waves in \emph{Region~B} marked in Fig.~\ref{fig:full_map}. The dominant emission pattern, \mpo{however, comes from the average ripple profile and is compatible} with the ripple scale-length.

Fig.~\ref{fig:full_map}f shows the averaged and smoothed profile of the $\Bz$ magnetic-field fluctuations taken along an oblique direction, as marked with a blue parallelogram in panel (c).
\mpo{High-intensity patches are spaced every $(2-3.5)\Lsi$,} consistent with $\lambda_{\rm rippl}$. Similar wave profiles are observed in $\Ex$ and $\Ey$ (not shown). \mpo{Short-wavelength fluctuations in $\Bz$ have associated electric field in the $x-y$ plane, approximately perpendicular to the wave vector, which suggests they are X-mode waves. Their spectra have cutoffs, indicated as white lines in Fig.~\ref{fig:fourier2}(a-b), that arise from the requirement that the waves be faster than the shock. While the peculiar structure of the precursor waves in our mildly relativistic shock results from shock rippling, the emission mechanism appears to be generic and corresponds to the well-known electron synchrotron maser.}

In addition to the dominant oblique component, the power spectrum of $\Bz$ oscillations in Fig.~\ref{fig:fourier2}(a) shows parallel waves with \mpo{$\kbzx \Lsi \simeq (10-30)$ that have an electric counterpart in $\Ey$, not in $\Ex$. The wakefield in $\Ex$
has a wave number $\kexx\Lsi\simeq 2$ and wide distribution in $\kexy$ reflecting the entire range of obliquity of the precursor waves. The wave number of the wakefield agrees with that expected in the standard electron SMI scenario, as estimated in equation~\ref{eq:exwavelenght}. It non-linearly couples to magnetic-field and density perturbations in the same wave band} (Figs.~\ref{fig:fourier2}a and~\ref{fig:fourier2}c).    

\subsubsection{Filamentation via parametric instability \label{sec:filam}}
\mpo{The Fourier spectrum of electron density in Fig.~\ref{fig:fourier2}(c) shows significant wave power at $k_y\Lsi\sim 10-30$ and $k_x\Lsi\sim 2-4$. The corresponding density perturbations are highlighted in Fig.~\ref{fig:full_map}(e). They form oblique filamentary structures whose transverse scale is a few $\Lse$.} We interpret these perturbations as result of the parametric filamentation instability \citep{kaw1973,drake1974} triggered when intense electromagnetic waves interact with the incoming upstream plasma. Similar filaments in density and magnetic field have been recently identified in high-resolution studies of ultrarelativistic magnetized pair shocks \citep{iwamoto2017,iwamoto2018,plotnikov2018} and electron-ion shocks \citep{iwamoto2019}. Their presence indicates coherence and self-focusing of the precursor waves in 2D systems. The filaments observed in pair plasmas largely retain their structure \mpo{during advection toward the shock. In electron-ion plasma, instead,} the filaments quickly merge to form long, ion-scale turbulent structures ahead of the shock. At our mildly-relativistic shock the filaments resemble those at pair shocks. They \mpo{are observable very far upstream, up to $x/\Lsi\sim 200$, but their structure is disrupted by the oblique waves, and their amplitude is only $\delta N_e/N_0\approx 0.1$. The corresponding spectral signature in the magnetic field is very weak (compare Fig.~\ref{fig:fourier2}(a)). Although at mildly relativistic shocks the precursor waves are less prominent than in the ultrarelativistic case, and the parametric instability is weakly driven, our high-resolution simulations can still detect coherent precursor-wave emission.}

\begin{figure}
\begin{center}
\includegraphics [width=0.99\linewidth] {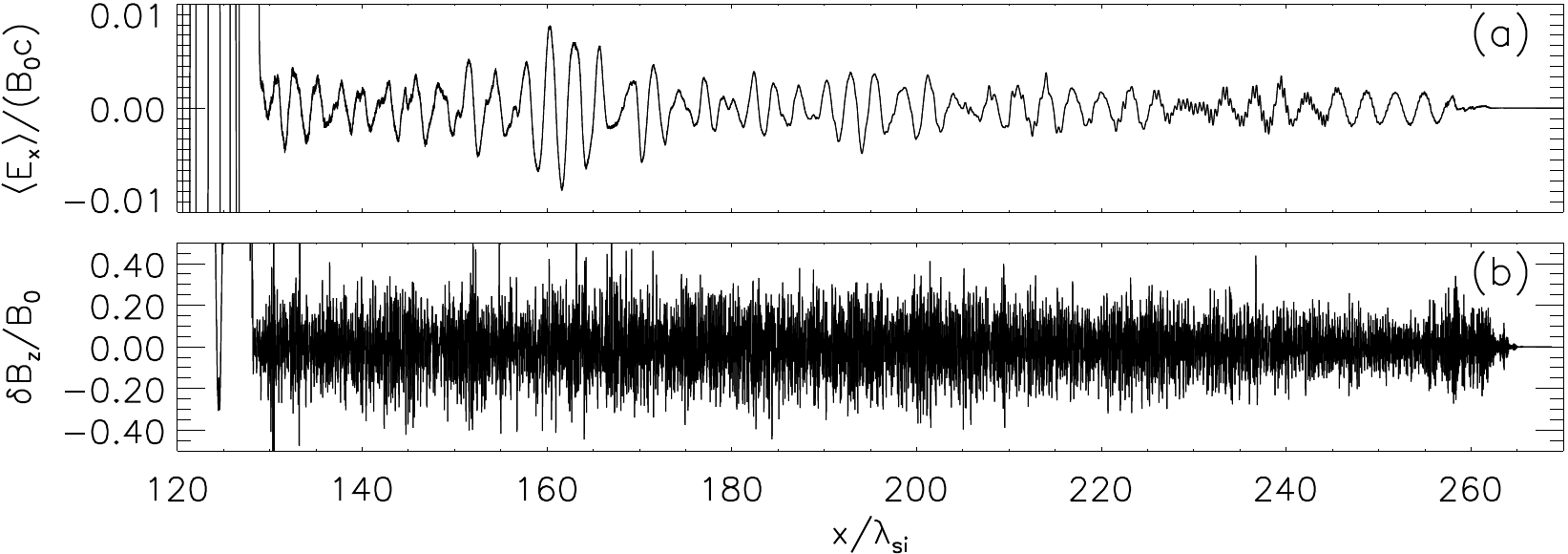}
\end{center}
\caption{\mpo{Normalized profiles along the shock normal at time $t\Wci = 84.3$ of $\langle\Ex\rangle$ averaged over $y$ (a) and $\delta\Bz$ measured in the middle of the box at $y/\Lsi=6$ (b).}}
\label{fig:profile}
\end{figure}

\begin{table}
\caption{Normalized precursor wave amplitudes, $\delta B/B_0$, and energy density, $\epsilon_{\rm pe}=\delta B^2/(2\mu_0\gamma_0N_em_ec^2)$, for 2D simulations with $\varphi_B=90\degree$ and $\varphi_B=0\degree$ and 1D simulation. For comparison we also list results of \citet{gallant1992}, marked "a", and \citet{iwamoto2019}, marked "b".
Here, $\delta B= \delta B_z$ for $\varphi_B=90\degree$ and 1D runs, and $\delta B= \sqrt{\delta\Bx^2+\delta\By^2+\delta\Bz^2}$ for $\varphi_B=0\degree$. The amplitudes are averaged in a region of length $\delta x=5\Lsi$ located $2\Lsi$ upstream of the shock.
The field fluctuations in \citet{gallant1992} were measured in the entire precursor region. \citet{iwamoto2019} calculated $\delta B$ and $\epsilon_{\rm pe}$ in a region about $70\Lsi$ in length and $32\Lsi$ upstream of the shock.}
%\label{table1}
\centering
\begin{tabular}{lcc}
\hline
\hline
\noalign{\smallskip}
  & $\delta B/B_0$ & $\epsilon_{\rm pe}$ \\
\noalign{\smallskip}
\hline
\noalign{\smallskip}
2D $i-e^-$, $\varphi_B=90\degree$ & $0.19\pm 0.01$  & $0.09\pm 0.005$ \\
2D $i-e^-$, $\varphi_B=0\degree$ & $0.15\pm 0.01$  & $0.07\pm 0.005$ \\
1D $i-e^-$  & $0.18\pm 0.01$  & $0.08\pm 0.005$ \\
\noalign{\smallskip}
\hline
\noalign{\smallskip}
1D$^{\rm a}$ $e^+\!\!\!-e^-$ & 0.46$^{+0.18}_{-0.12}$ & 0.53$^{+0.22}_{-0.15}$ \\
2D$^{\rm b}$ $e^+\!\!\!-e^-$ & $0.064\pm 0.031$  & $0.010\pm 0.005$  \\
1D$^{\rm b}$ $e^+\!\!\!-e^-$ & $0.10\pm 0.01$  & $0.025\pm 0.005$ \\
2D$^{\rm b}$ $i-e^-$ & $0.50\pm 0.10$  & $0.65\pm 0.25$ \\
1D$^{\rm b}$ $i-e^-$ &  $0.75\pm 0.09$  & $1.4\pm 0.4$ \\
\noalign{\smallskip}
\hline
\end{tabular}
\label{table1}
   \end{table}

\subsubsection{Precursor wave amplitudes \label{precursor_waves}}
\mpo{Fig.~\ref{fig:profile} shows profiles in the precursor region of $\delta\Bz$, taken at $y/\Lsi=6$, and $\langle\Ex\rangle$ $y$-averaged to filter out the contribution of the oblique large-amplitude precursor waves. 
These profiles can be compared to corresponding profiles in 1D simulation that we describe in Appendix~\ref{app:1d}.}

\mpo{The magnetic-field fluctuation amplitude \jn{normalized to the upstream field strength, $\delta B/B_0$}, and  energy density \jn{normalized to the upstream \emph{electron} kinetic energy}, $\epsilon_{\rm pe}=\delta B^2/(2\mu_0\gamma_0N_em_ec^2)$, are listed in Table~\ref{table1} in comparison with results of other studies. 
For out-of-plane magnetic field the only relevant polarization is $\delta B=\delta\Bz$.}
The amplitudes are averaged in the region of $x/\Lsi=129-134$, located about $2\Lsi$ from the shock. \mpo{Our 1D test simulation is described in detail in Appendix~\ref{app:1d}. The amplitude of the precursor waves is comparable in 2D and 1D simulations. In 2D, the average wave amplitude out to $x/\Lsi\approx 230$ is even larger than the strongly coherent oscillations at the tip of the precursor, that were generated in the early linear phase (see Fig.~\ref{fig:full_map200}). Fig.~\ref{fig:zoom_1d} demonstrates that in 1D the waves closer to the shock are weaker, possibly due to heating of the electrons that suppresses higher harmonics in the electron SMI \citep{amato2006}. In a 2D simulation the same should happen, and inhomogeneities at the shock may further cause a reduction of wave coherency. But we observe that shock rippling is highly organized and produces a semi-coherent, modulated train of oblique precursor waves. Thus, instead of being destructive, the ripples amplify the precursor-wave amplitude.}

\mpo{The magnetic-field amplitude can be compared to that observed at ultra-relativistic shocks. Since electron magnetization is a relevant parameter, in Table~\ref{table1} we list available results for shocks with $\sigma_e=5$. As expected, both $\delta B/B_0$ and $\epsilon_{\rm pe}$ are smaller than those in 1D simulations of pair shocks \citep{gallant1992}, but they are much larger than those obtained in the high-resolution 1D and 2D simulations of pair shocks by \citet{iwamoto2019}.
The wave energy is much smaller than that at ion-electron shocks with $\gamma_0=40$ \citep{iwamoto2019}, which both in 1D and 2D exceeds that in pair plasmas by almost two orders of magnitude, and the 2D amplitude is close to that in 1D. The high wave intensity at high-$\gamma_{\rm sh}$ ion-electron shocks was attributed to the so-called positive feedback, in which incoming electrons accelerated by the wakefield cause enhanced precursor-wave emission, that in turn amplifies the wakefield. At the mildly relativistic shocks described here, the wakefield does not reach a very high amplitude (see Section~\ref{sec:wakefield}), and the positive feedback is not effective. However, electromagnetic precursor wave amplification up to the 1D level is achieved through shock rippling.}
%facilitating the electron-ion coupling

\begin{figure}
\begin{center}
\includegraphics [width=0.99\linewidth] {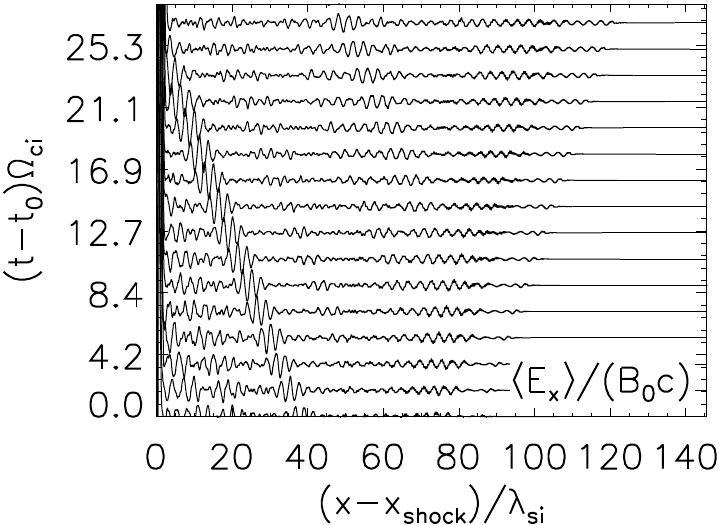}
\end{center}
\caption{Stack plot of the averaged wakefield \mpo{profile, $\langle E_x\rangle/(B_0c)$, between time $t_0\Wci=56.2$ and $t_{\rm max}\Wci=84.7$. The shock is always located at the left edge of the panel.}}
\label{fig:stacked_profile}
\end{figure}

\subsubsection{Wakefield waves \label{sec:wakefield}}

As noted before, the ponderomotive force on the upstream plasma leads to longitudinal plasma \mpo{motions and associated electrostatic wakefield, whose average amplitude does not exceed $\langle\Ex\rangle/(B_0c)\sim 0.01$. The weakness of the wakefield reflects the relatively low amplitude of the precursor waves, that can be expressed through the so-called strength parameter $a$ that can be \jn{estimated} as \citep{iwamoto2017}
\begin{equation}
    a\simeq\gamma_0\sqrt{\sigma_e}\frac{\omega_{\mathrm{pe}}}{\omega}\frac{\delta B}{\Bo}.
\label{eq:a}
\end{equation}
Here $\omega$ is the wave frequency. If $a \gtrsim 1$, the precursor waves can generate intense wakefield  \citep{kuramitsu2008}.
Fig.~\ref{fig:fourier2}(a) indicates typical wave numbers in the range $k\approx (15-40)\Lsi^{-1}\simeq (2.1-5.7)\Lse^{-1}$. The dispersion relation in equation~\ref{eq:xmodedisp_approx_fin} then gives for the wave frequency $\omega\approx (2.5-6)\,\omega_{pe}$.
The average magnetic-field amplitude is $\delta\Bz/B_0\simeq 0.19$ (see Table~\ref{table1}), and all together we find for the strength parameter
\begin{equation}
    a\approx (0.14-0.35) . 
\end{equation}
The corresponding amplitude of the wakefield can be estimated following \citet{hoshino2008}, using $\xi=1/2$ as for a linearly polarized wave:
\begin{equation}
    \frac{\langle\Ex\rangle}{B_0c}\simeq \frac{\xi a^2}{\sqrt{1+\xi a^2}}\,\frac{1}{\sqrt{\sigma_e}\gamma_0}\approx (2-13)\cdot 10^{-3},
\label{eq:wakefield}
\end{equation}
in agreement with our simulation result.}

Shock rippling causes enhanced emission of the precursor waves at oblique angles which in turn produces oblique Langmuir waves. \mpo{They are averaged out of the wakefield profile shown in Fig.~\ref{fig:profile}(a), and so the local wave amplitude may be} \jn{much larger than the mean} amplitude. In fact, from time $t\Wci\sim 40$ on, when the oblique precursor-wave structure is well established, episodes of stronger semi-coherent wave emission from the shock lead to stronger wakefield in the near-upstream region, \mpo{that non-linearly evolves.}

Fig.~\ref{fig:stacked_profile} shows a stack plot of averaged wakefield profiles \mpo{for a time period $28.1\,\Wci^{-1}$, starting at $t_0\Wci=56.2$. The profiles are given in shock-centered coordinates, $x-x_\mathrm{shock}$.
Far upstream of the shock, the electrostatic waves propagate away from the shock, but within $70\,\Lsi$ of the shock the wakefield on average moves back toward the shock. }
\jn{We interpret} this {downstream-directed motion of the wakefield} 
\jn{as result of Forward Raman Scattering (FRS)}
operating at our mildly relativistic shock. 

The ponderomotive force is proportional to the gradient of the wave pressure and can also act inside the precursor, if the electromagnetic waves are modulated \citep{hoshino2008}. 
The enhanced emission of the precursor waves through shock rippling amplifies the waves and \jn{triggers the nonlinear FRS. In this process} 
the scattered electromagnetic waves successively decay into other electromagnetic waves and Langmuir waves.
{As the frequency of the scattered wave is lower than that of the pump wave, broadband precursor wave spectra extending from the initial $\omega'\gtrsim\Wce$ down to $\omega'>\wpe$ are generated. Similarly, broadband Langmuir waves are produced with $k'_L<\wpe/c$ and $\omega'_L\simeq\wpe$} \citep{hoshino2008}.
In the upstream plasma rest frame the electromagnetic and Langmuir waves \mpo{all} have phase velocities in the upstream direction, \mpo{but} in the simulation frame part of these waves move toward downstream. 

\mpo{The presence of large-amplitude wakefield propagating toward the shock is of importance for electron energization upstream of the shock. We will show in Paper II that the waves can scatter electrons and boost them toward the shock, contributing to ion-to-electron energy transfer in the precursor region. }

\section{Comparison with the in-plane setup} \label{sec:in-plane}
In this section we compare \jn{the electromagnetic structure of} a mildly relativistic shock with upstream magnetic field lying in the plane of the simulation, $\boldsymbol{B_0}=\By\boldsymbol{\hat{y}}$, \jn{thus $\varphi=0\degree$}, \mpo{with that for out-of-plane field discussed in Section~\ref{sec:out-of-plane}. 
With in-plane magnetic field the shock quickly acquires its steady-state form, and so we show the structure and discuss its properties only at time $t\Wci= 84.3$.}

\begin{figure}
\centering
\includegraphics [width=0.99\linewidth] {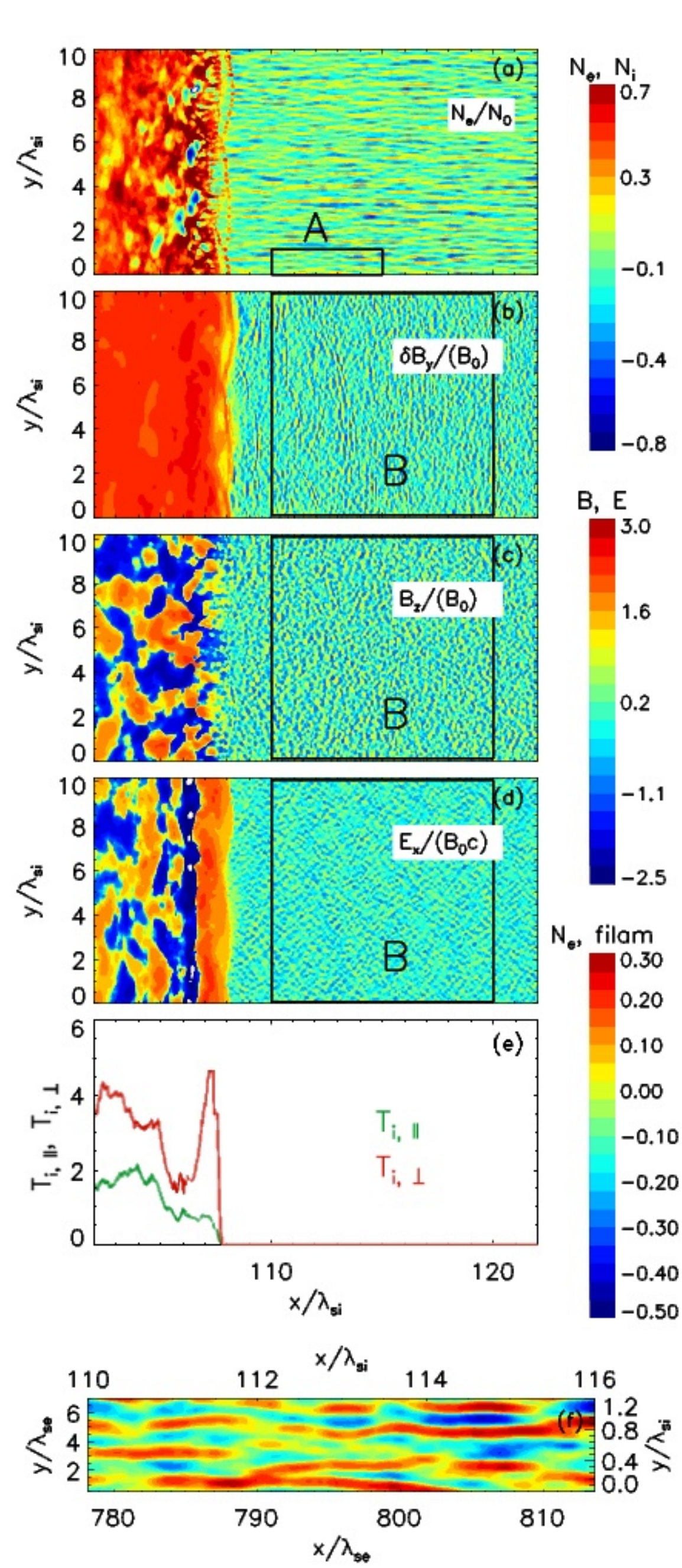}
\caption{Normalized electron density (a), field fluctuations in $\By$ (b), $\Bz$ (c), and $\Ex$ (d), \jn{as well as the averaged profile of the ion temperature parallel and perpendicular to the mean magnetic field (e)}, at the final stage of the simulation, at $t\Wci = 84.3$. Logarithmic scaling is applied as in Fig.~\ref{fig:structure_early}. Panel (f) shows a close-up of Region~A in the electron density plot (a). For region~B we show Fourier spectra in Fig.~\ref{fig:fourier_B_ip}.}
\label{fig:maps_inplane}
\end{figure}

\subsection{Shock front and downstream turbulence}
\mpo{To be noted from Fig.~\ref{fig:maps_inplane} is the lower shock speed, $v=0.41c$, compared to the out-of-plane case, that is implied by its location at $x/\Lsi\simeq 108$.} The density compression is %\jn{$N_{\rm d}/N_{\rm u}\simeq 3.2$}
\jn{$\kappa\simeq 3.2$}. \mpo{Both are consistent with a shock in} plasma with three degrees of freedom and adiabatic index $\Gamma ={4/3}$ \citep{plotnikov2018}. \mpo{Ion gyration at the shock happens in the $x-z$ plane, which suppresses the gyration-driven rippling mode seen with out-of-plane magnetic field.}

Fluctuations in density and electromagnetic field can be observed together with corrugations in the shock surface, that develop very early in the simulation and quickly evolve into a large-scale rippling mode with $\lambda_{\rm rippl}\simeq 5\Lsi$ \mpo{and amplitude $\lesssim \Lsi$.} They propagate along the mean magnetic field and are most likely driven by the anisotropy in the ion temperature that results from ion reflection from the shock \mpo{and was geometrically suppressed in the out-of-plane simulation.} \jn{Fig.~\ref{fig:maps_inplane}(e) demonstrates that at the shock $T_{\mathrm{i}\, \perp} \gg T_{\mathrm{i}\, \parallel}$} \mpo{with respect to the mean magnetic field direction, which triggers the Alfv\'en Ion Cyclotron (AIC) instability that is known to produce} ripples in low-Mach-number shocks \citep[e.g.][]{winske1988,2014PhPl...21b2102U} and can generate magnetic-field fluctuations at the front of relativistic pair shocks \citep{iwamoto2018}. Shock-front corrugations are also a source of downstream turbulence, through the mechanism of the vorticity generation via a process similar to the Richtmyer-Meshkov instability \citep[e.g.][]{2011ApJ...726...62M,2014MNRAS.439.3490M}.

\begin{figure}
\centering
\includegraphics [width=0.99\linewidth] {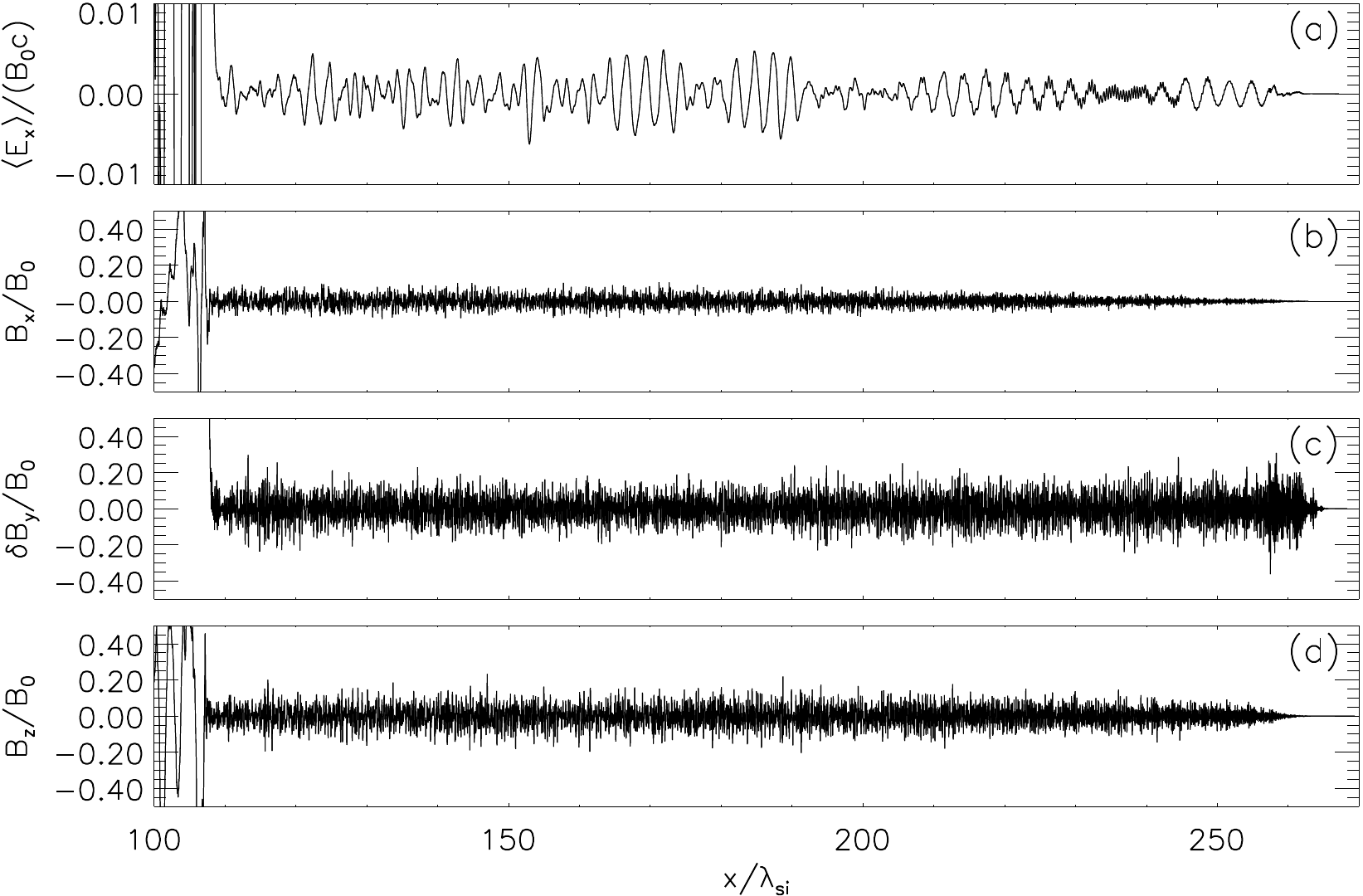}
\caption{Profiles along the shock normal of normalized \mpo{field components at time $t\Wci = 84.3$ for in-plane upstream magnetic field configuration. We averaged $\langle\Ex\rangle$ over $y$ (a), whereas $\Bx, \delta\By=(\By-B_0)$, and $\Bz$ are measured} in the middle of the box along $y/\Lsi=6$ (b-d) (see also Fig.~\ref{fig:profile}).
\label{fig:shock_str_z_ip}}
\end{figure}

\begin{figure}
\centering
\includegraphics [width=0.99\linewidth] {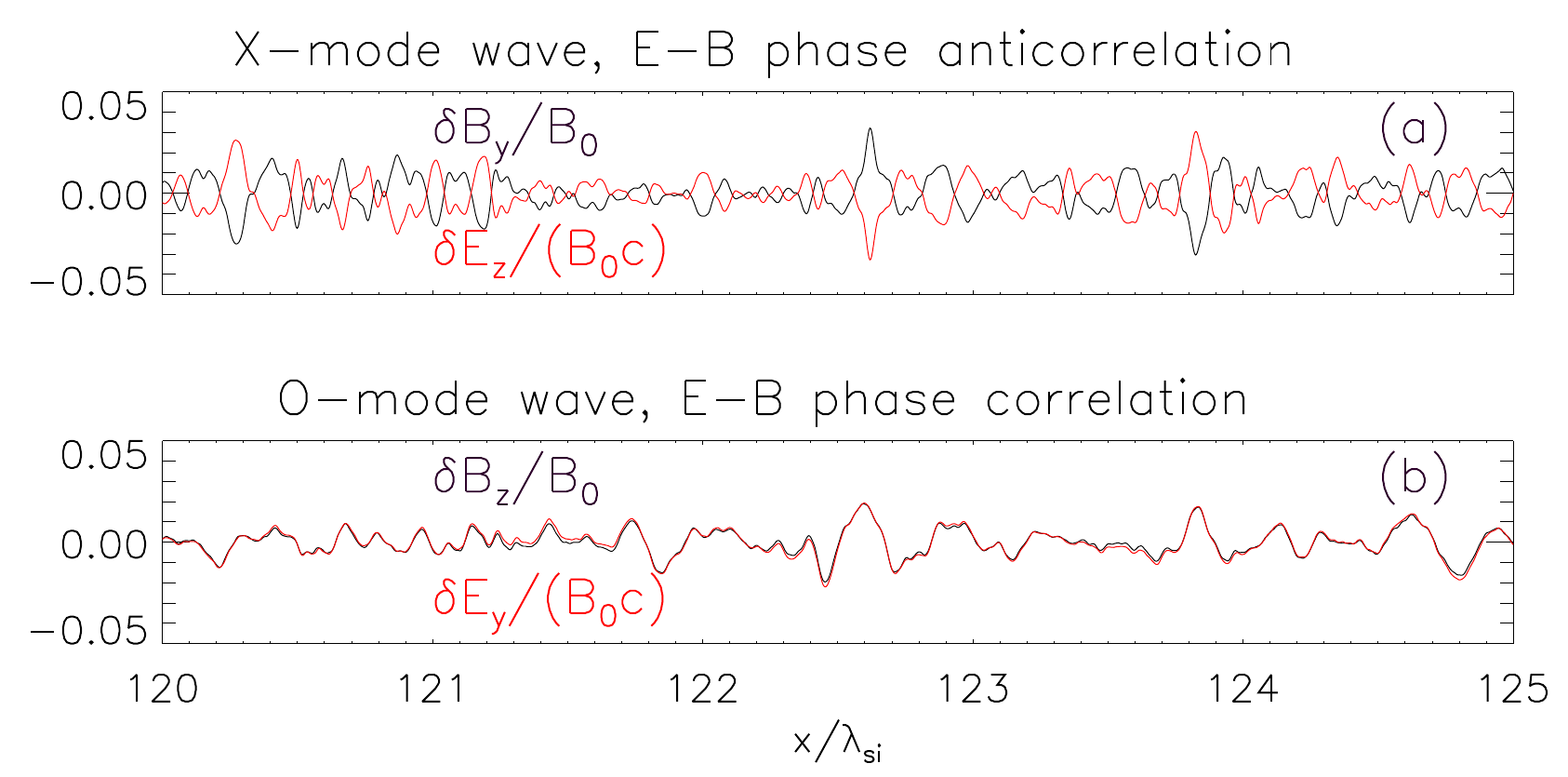}
\caption{Enlarged view of the region $x/\Lsi=120-125$ \mpo{with magnetic-field and the electric-field profiles plotted in black and red, respectively, and normalized as in Fig.~\ref{fig:shock_str_z_ip}. The phase (anti-)correlation between $y$ and $z$ field components indicate X-mode (a) and O-mode (b) waves.}}
\label{fig:profiles_correl}
\end{figure}

\subsection{Upstream waves}
\label{sec:ip_waves}
Short-scale precursor waves and large-scale electrostatic wakefield are evident in Fig.~\ref{fig:maps_inplane} and in the profiles covering the entire upstream region at time $t\Wci = 84.3$, that we plot in
Fig.~\ref{fig:shock_str_z_ip} in the same manner as earlier for 2D out-of-plane simulation and in Appendix~\ref{app:1d} for a 1D test run. 
Fig.~\ref{fig:profiles_correl} shows an enlarged view of the region $x/\Lsi=(120-125)$, presenting also electric-field fluctuations. \mpo{Electromagnetic waves with $\By$ fluctuations parallel to the large-scale magnetic field, $\boldsymbol{B_0}=\By\boldsymbol{\hat{y}}$, and anticorrelated with $\Ez$ fluctuations are X-mode waves. 
Correlated $\Bz$ and $\Ey$ fluctuations are O-mode waves. 
%$\boldsymbol{k}=k_x\boldsymbol{\hat{x}}$. 
Oscillations in $\Bx$ result from an oblique propagation of the X-mode wave, as we discuss below.}

The magnetic-field fluctuation amplitudes are listed in Table~\ref{table1}. 
The time evolution of the field amplitudes is shown in Fig.~\ref{fig:ampl_a_oop} in Appendix~\ref{app:1d}. We list the total amplitude of the magnetic field oscillations, $\delta B=\sqrt{\delta\Bx^2+\delta\By^2+\delta\Bz^2}$, not differentiating between the X-mode and O-mode waves. The amplitude of the X-mode wave, $\delta B_X/B_0=\sqrt{\delta\Bx^2+\delta\By^2}/B_0\simeq 0.12$, is larger than the amplitude of the O-mode wave, $\delta B_O/B_0=\delta\Bz/B_0\simeq 0.08$. The total precursor wave amplitude, $\delta B/B_0\simeq 0.15$, is slightly smaller than the amplitudes obtained in our 2D run with \jn{$\varphi_{\rm B}=90\degree$} and in the 1D simulation. 

\mpo{With out-of-plane magnetic field strong shock ripples increase the precursor-wave amplitude to the level observed in 1D simulation. With in-plane field we see a} similar amplification. \mpo{Fig.~\ref{fig:maps_inplane}(b) shows the emission of precursor waves in bunches that correspond to rippling features at the shock.} The rippling driven by the AIC instability is relatively weak and \mpo{cannot fully compensate losses in coherency due to random fluctuations in shock structure at the shock surface and the thermal damping of the waves. The precursor wave amplitude is large enough, though, to induce the wakefield and \jn{thus} accelerate and heat particles}. Fig.~\ref{fig:ampl_a_oop} shows a roughly constant wave amplitude throughout the 2D simulation with the in-plane setup, which is due to an early formation of the shock ripples.    

\mpo{The linear theory of the SMI predicts X-mode emission at a level above that of O-mode waves \citep{melrose1984, lee1980, wu1979}. In earlier 2D simulations of ultrarelativistic pair-plasma shocks with in-plane magnetic field the O-mode energy was observed to exceed that in X modes at very small electron magnetizations, $\sigma_e\lesssim 10^{-2}$ \citep{iwamoto2018}, which was attributed to mode conversion from X to O modes at the turbulent shock transition. Early in} the shock evolution, charged particles gyrate in the $x-z$ plane and emit X-mode waves with $\delta\By$. As plasma instabilities generate fluctuations in $\Bz$ of amplitude comparable to $B_0$, the net magnetic field undulates in the $y-z$ plane, and the X-mode waves carry both $\delta\By$ and $\delta\Bz$ fluctuations. \mpo{Upon transmission to the upstream region the $\delta\Bz$ components may be converted into O-mode waves.}
Indeed, Fig.~\ref{fig:shock_str_z_ip} shows the tip of the $\Bz$ turbulence behind that of $\By$, indicating that the O-mode waves are \mpo{produced a bit later than} the X-mode waves, after the shock front has developed substantial turbulence. The slightly smaller amplitude of the O-mode waves with respect to the X-mode waves reflects the moderate level of $\Bz$ fluctuations in the shock front, $\delta\Bz/B_0\approx 1$ (compare results for $\sigma_e\gtrsim 10^{-2}$ in \citet{iwamoto2018}).

\begin{figure}
\centering
\includegraphics[width=0.91\linewidth]{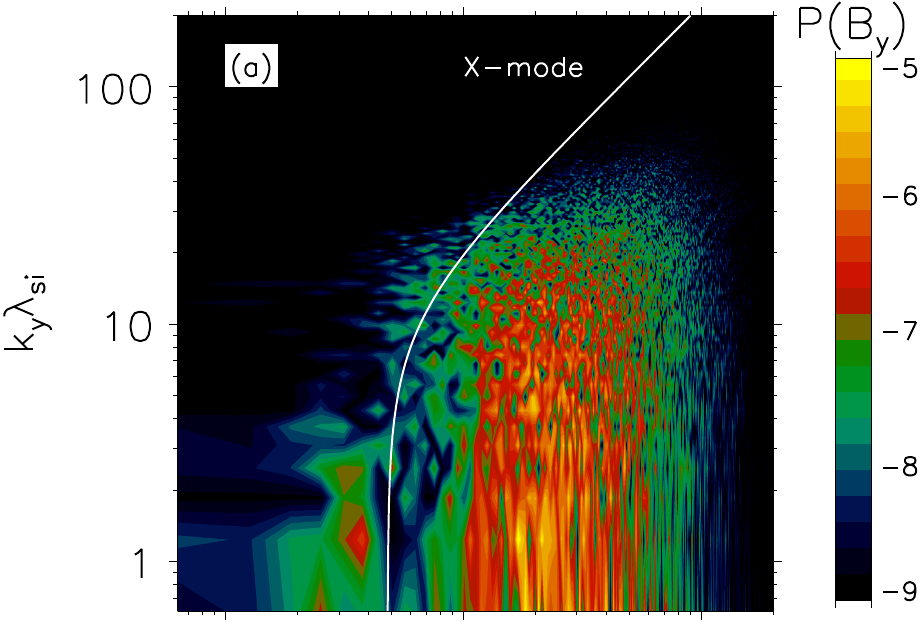} 
\includegraphics[width=0.91\linewidth]{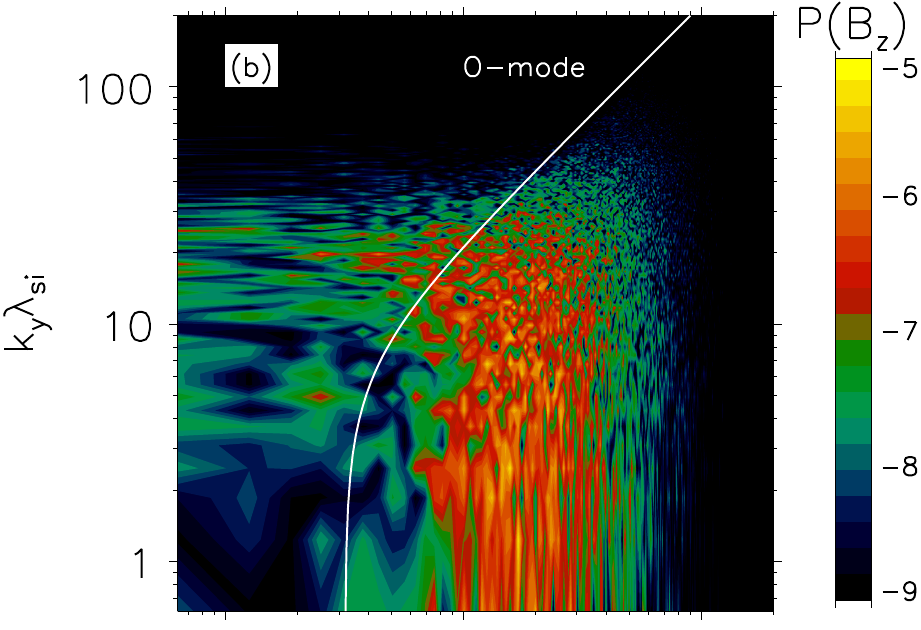}
\includegraphics[width=0.91\linewidth]{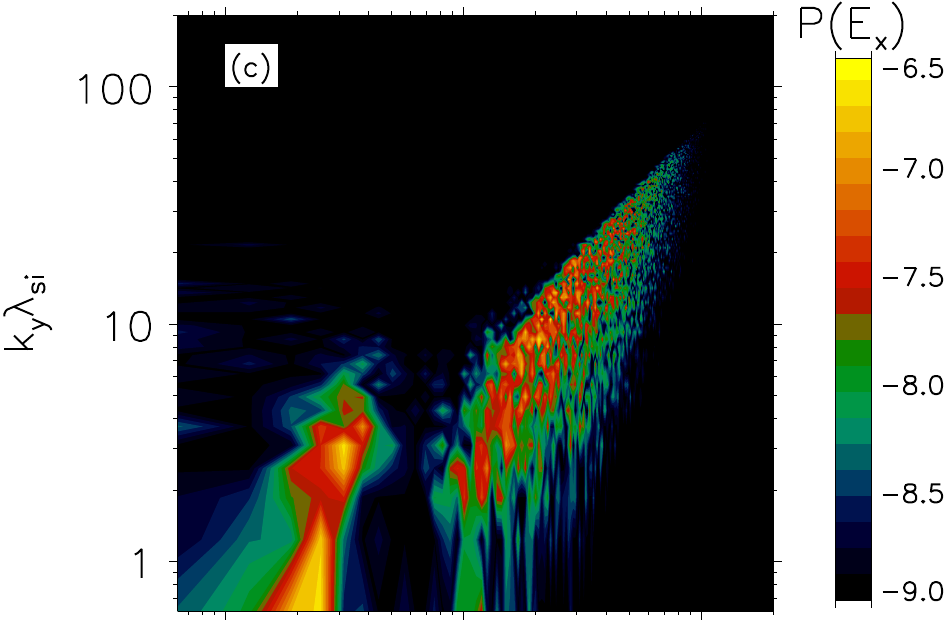}
\includegraphics[width=0.91\linewidth]{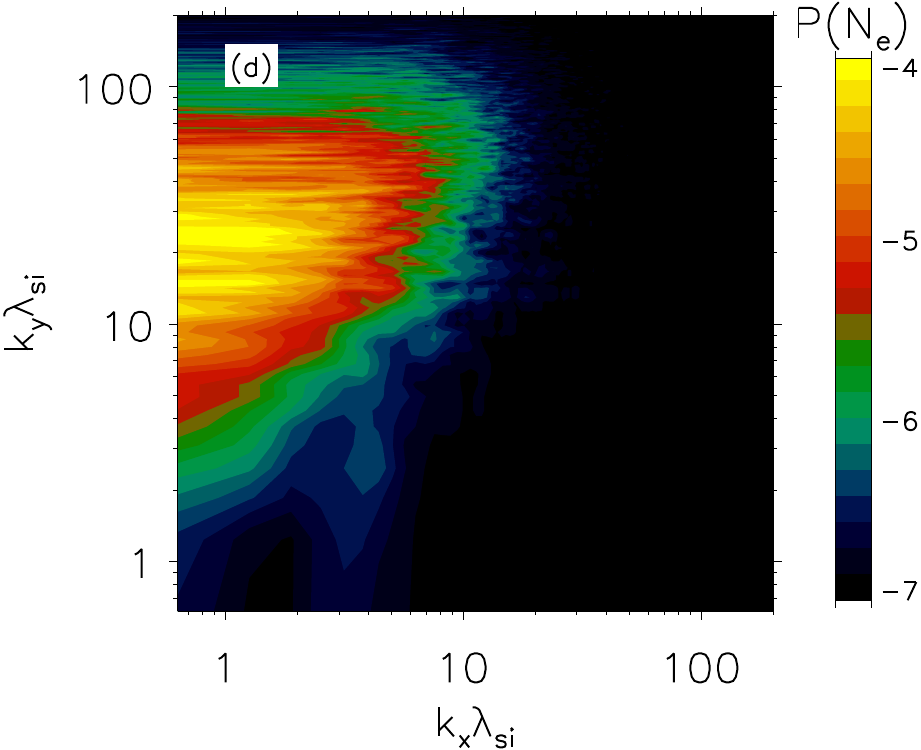}
\caption{Fourier power spectra for $\By$ (a), $\Bz$ (b), $\Ex$ (c), and the electron density (d) for Region B at $x/\Lsi=110-120$ at time $t\Wci = 84.3$ (see Fig.~\ref{fig:maps_inplane}). The white lines show the cutoff for the X-mode (a) and the O-mode (b) waves.}
\label{fig:fourier_B_ip}
\end{figure}

Fig.~\ref{fig:fourier_B_ip} shows Fourier power spectra of $\By$, $\Bz$, $\Ex$, and the electron density in \emph{Region~B} marked in Fig.~\ref{fig:maps_inplane}. \mpo{Most of the wave power in $\delta\By$ and $\delta\Bz$ is located to the right of the theoretical cutoff calculated in Appendix~\ref{app:lintheory}, and the waves are faster than the shock. The wave vector range of the precursor waves is similar to that in the out-of-plane setup, and an oblique component is likewise present. Coherent precursor waves are persistent in mildly relativistic shocks, albeit with smaller amplitude than in the ultrarelativistic regime.} 

\mpo{Also with $\boldsymbol{B_0}$ in the simulation plane we observe transverse density filaments upstream of the shock, whose amplitude, $\delta N_e/N_0\approx 0.5$, is much larger than in the out-of-plane simulation. They are better aligned with the $x$-direction, though. It appears that the parametric filamentation instability is not  affected by the weak ripples at the shock.}

\mpo{The wakefield is evident upstream of the shock, as for $\varphi_{\rm B}=90\degree$. Fig.~\ref{fig:fourier_B_ip}(c) shows a signal at $\kexx\simeq 2.5\Lsi^{-1}$ that is marginally consistent with equation~\ref{eq:exwavelenght}. The wave power is slightly less than that observed with out-of plane magnetic field. As in Section~\ref{sec:wakefield} we take the typical wavenumber of the precursor waves, $k\approx 2.8\Lse^{-1}$, and the typical frequency of the dominant X-mode waves, $\omega\approx 3.2\,\omega_{pe}$, and add the contributions from the X- and O-modes to estimate the strength parameter and calculate the amplitude of the wakefield (equatio~\ref{eq:wakefield}),
\begin{equation}
    a\simeq 0.21 \qquad \frac{\langle\Ex\rangle}{B_0c}\approx 5\times 10^{-3}.
\end{equation}
Therefore, the wakefield should exert similar effects on the upstream plasma in simulations with in-plane and out-of-plane magnetic field.}

\section{Summary and conclusions} \label{sec:summary}
\mpo{This is the first of two articles in which 
we investigate mildly-relativistic magnetized shocks in electron-ion plasma. In this paper we explore  with PIC simulations the electromagnetic shock structure and
the production of plasma instabilities and waves. Paper II shall be devoted to particle acceleration,  heating, and the energy transfer from ions to electrons downstream of the shock. }

\mpo{Our high-resolution studies show that the SMI operates at mildly relativistic shocks as theoretically predicted and produces coherent emission of upstream-propagating electromagnetic waves. The waves are substantially weaker than at ultra-relativistic shocks \citep{iwamoto2017,iwamoto2018}, but reach a persistent level that is similar in 2D and 1D simulations. In 2D shock corrugation provides wave amplification that compensates other destructive 2D effects. Shock ripples appear for both in-plane and out-of-plane mean magnetic field, but their generation mechanism differs -- modulation of ion gyration \citep{burgess2007} for $\varphi_B=90\degree$ and the AIC temperature-anisotropy instability with $\varphi_B=0\degree$. 
In both cases} the ripples heavily influence the upstream plasma and the structure of downstream turbulence. 

\mpo{For out-of-plane mean magnetic field the precursor waves are of the X-mode type. Both the emission direction about $40\deg$ off the shock normal and the wave amplification are caused by the shock ripples. With in-plane magnetic field the AIC-generated shock-front corrugations have a slightly lower amplitude, and the waves propagate mostly along the shock normal. Magnetic turbulence at the shock causes part of the precursor waves to be O-mode waves. For both magnetic-field orientations we observe in the upstream plasma the electrostatic Langmuir modes -- the wakefields -- and the density filaments that result from the parametric filamentation instability. Except for brief periods, their average amplitude is moderate and smaller than at ultrarelativistic shocks.} 

\mpo{The important role of shock rippling has not been demonstrated so far for relativistic shocks. At perpendicular high-Lorentz-factor shocks \citet{sironi2013} found shock corrugations consistent with \citet{burgess2007} only in a limited range of plasma magnetization, 
$3\times 10^{-4}\lesssim\sigma \lesssim 10^{-1}$, probably on account of electron heating at Weibel filaments for $\sigma \leq 10^{-4}$ and in the SMI-mediated precursor for $\sigma > 0.1$. The ripples at mildly relativistic shocks may be similarly suppressed at low magnetizations \jn{due to} the Weibel instability \jn{that still operates in this regime} \citep[e.g.,][]{Kato2010}, but at $\sigma> 0.1$ one does not expect precursor wave emission stronger than for the $\sigma=0.1$ analysed here \citep[see][]{iwamoto2019}, and rippling may persistently develop.} {The same should apply to AIC-instability-generated corrugations.}

\mpo{Our 2D simulations show intense coherent precursor waves generated by the SMI {irrespective of the} magnetic-field configuration. One should expect strong precursor waves also in 3D, that are a mixture of X-mode and O-mode waves. The intensity of the ordinary mode is difficult to estimate, because these waves arise from local variations in the gyration direction at the shock front that may be less coherent in 3D than in the in-plane 2D simulations, possibly leading to weaker O-mode emission \citep{iwamoto2018}. 
\citet{plotnikov2019} showed for {ultra}-relativistic shocks in pair plasma that at high $\sigma_e\gtrsim 1$ ($\sigma_e=5.1$ in this study) the physics of precursor-wave emission
in 3D {is} better represented {with} the out-of-plane 2D {model.} 
This result will likely hold in electron-ion plasma, since the SMI mechanism operates as in pair plasma. As we demonstrated, at mildly relativistic shocks the precursor-wave strength and structure is significantly affected by shock rippling. Rippling along the lines of \citet{burgess2007} requires a suppression of fluctuations parallel to the mean magnetic field that might be difficult to achieve in 3D, but ripples generated through the AIC instability amplify the precursor waves to comparable amplitudes, and so precursor-wave amplification may be expected in 3D as well.}

\section*{Acknowledgements}
\jnr{J.N. acknowledges inspiring discussions with Marek Sikora.}  
This work has been supported by Narodowe Centrum Nauki through research projects DEC-2013/10/E/ST9/00662 (A.L., J.N., O.K.), UMO-2016/22/E/ST9/00061 (O.K.) and 2019/33/B/ST9/02569 (J.N.). This research was supported by PLGrid Infrastructure. Numerical experiments were conducted on the Prometheus system at ACC Cyfronet AGH. This work was supported by JSPS-PAN Bilateral Joint Research Project Grant Number 180500000671.
Part of the numerical work was conducted on resources provided by the North-German Supercomputing Alliance (HLRN) under projects bbp00003, bbp00014, and bbp00033.

\section*{Data availability}
The data underlying this article will be shared on reasonable request to the corresponding author.

\appendix 
\section{Numerical convergence tests}
\label{app:conv_test}

\begin{figure}
\begin{center}
\includegraphics[width=1.0\linewidth]{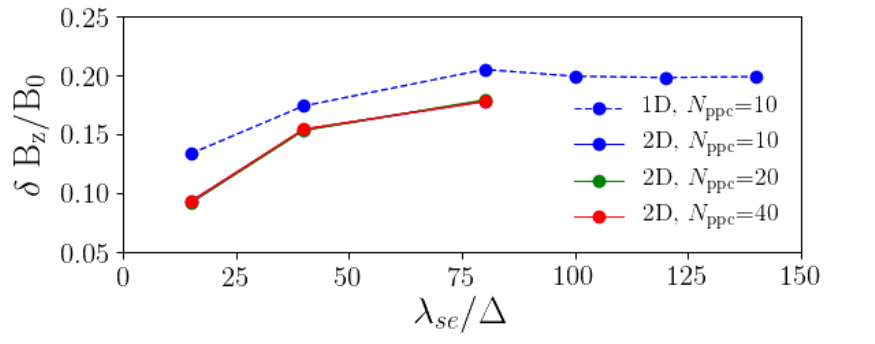}
\end{center}
\caption{Normalized \mpo{$\mathrm{rms}(\delta\Bz)$ amplitudes measured as function of $\Lse$ at $t\Wci\simeq 22.5$ about $2\Lsi$ upstream from the shock in small test simulations. The solid blue, green, and red lines for 2D runs with different $N_\mathrm{ppc}$ overlap.}}
\label{fig:convergence_summary}
\end{figure}

\mpo{\citet{iwamoto2017} demonstrated that simulations of precursor waves at magnetized superluminal shocks require very high resolution. 
We performed 2D test simulations with out-of-plane magnetic field, $\varphi_{\rm B}=90\degree$, to 
probe the impact of grid resolution, $\Lse/\Delta$, and the number of particles per cell, $N_{\rm ppc}$. The transverse size of the numerical grid was reduced to $L_y\simeq 3\Lsi$. 
In Fig.~\ref{fig:convergence_summary} we compare the normalized wave amplitudes at time $t\Wci\simeq 22.5$ \jn{(solid lines)}. The amplitudes are averaged in a region located about $2\Lsi$ upstream of the shock. The number of particles per cell does not influence the amplitude of the precursor waves, and we can choose $N_{\rm ppc} = 10$ for our 2D large-scale runs.}

\mpo{The wave amplitudes appear to increase for larger $\Lse/\Delta$. To explore the behavior beyond the maximum resolution tested with 2D simulations, $\Lse = 80\Delta$, we performed 1D test runs, extending the probed resolution up to $140\Delta$. 
The 1D tests indicate saturation of the wave amplitude at the resolution $\Lse = 80\Delta$, which we then used in our production runs. The amplitudes are generally slightly larger in 1D simulations than in 2D runs.}

\section{Linear theory of the synchrotron maser instability in electron-ion plasmas} \label{app:lintheory} 

\begin{figure}
\begin{center}
\includegraphics[width=0.99\linewidth]{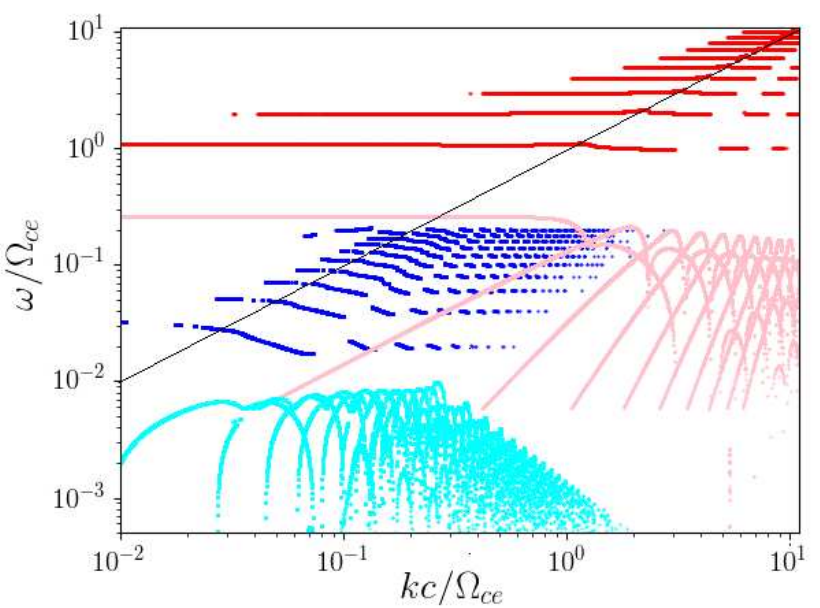}
\end{center}
\caption{Dispersion relation for X-mode waves \mpo{produced by cold gyrating electrons and ions in plasma} with electron magnetization $\sigma_e=5.1$. The oscillation frequency, \mpo{$\omega_r$, and growth rate, $\omega_i$, for the electron SMI are shown with red and pink} dotted lines, respectively. 
\al{Blue and cyan dots correspondingly represent the ion maser instability. The black line shows the lightwave dispersion relation, $\omega=kc$.}}
\label{fig:matsukyio_plot}
\end{figure}

Fig.~\ref{fig:matsukyio_plot} shows the X-mode dispersion relation for relativistic magnetized electron-ion plasmas with parameters used in this study. The dispersion relation was derived following \citet{hoshino1991}, but assuming that both electrons and ions form \emph{cold} rings in momentum space while they gyrate about the magnetic-field lines with $\gamma\simeq 3.3$. 
\citet{hoshino1991} treated electrons as a hot background fluid with temperature $\gamma_e m_e c^2$ when deriving the dispersion relation of the ion-generated SMI. The ring distribution that we use provides the same effective perpendicular temperature. In Fig.~\ref{fig:matsukyio_plot} \mpo{we show} the real part of frequency, $\omega_r$, and the imaginary part, $\omega_i$, \mpo{for gyrating electrons and ions with zero bulk (drift) velocity. The $k$ vector is perpendicular to the magnetic field. The emission spectrum has a clear harmonic structure for both the electron-generated SMI
($\omega_r / \Omega_{ce} \gtrsim 1$) and the ion-generated SMI ($\omega_r / \Omega_{ce} \lesssim 0.1$). We considered the first \jn{ten harmonics}  of the Bessel function when calculating the dispersion relation.}

For the electron SMI, the growth rate of the fundamental mode is comparable to that of the higher harmonics. The phase velocity of the \mpo{waves is about $c$ at the wavenumber of maximum growth. Note that the group velocity is small except near $\omega=kc$.}
{The frequencies and growth rates of the unstable ion-modes are lower than the electron ones by approximately the ion-to-electron mass ratio, and the growth rate slightly increases with higher harmonic number. One should thus expect two stages of the SMI to occur -- first the electron SMI, then the ion maser instability. However, the}
\jn{ions emit mainly left-handed elliptically polarized magnetosonic waves that are} mostly sub-luminous, the phase velocities of the modes are \jn{$\omega/kc\lesssim v_{\rm sh}$, so that} the waves emitted at the shock cannot outrun it to reach the precursor.

\mpo{The polarization of the electron SMI waves is that of X modes, and after transmission to the upstream medium the waves would propagate with the standard dispersion relation of X-mode waves that reads}
\begin{equation}
\eta'^2=1 - \frac{\omega_{pe}^2}{\omega'^2 - \sigma_e \,\omega_{pe}^2} \ ,
\label{eq:xmodedisp_ups}
\end{equation}
where the prime denotes quantities measured in the upstream plasma frame and $\eta'\equiv ck'/\omega'$ is the refraction index.  
Equation~\ref{eq:xmodedisp_ups} includes only the electron contribution and can be used for $\omega' \ge \omega_{pe} \sqrt{1+\sigma_e}$ \citep{hoshino1991}. 
\mpo{Following \citet{iwamoto2017} we} estimate the theoretical cutoff wave number, above which the waves emitted at the shock can escape upstream.
\mpo{Lorentz transforming equation \ref{eq:xmodedisp_ups}} one obtains the dispersion relation in the simulation rest frame: 
\begin{equation}
\frac{\omega^2}{\omega_{pe}^2}=\frac{\sigma_e}{\gamma^2 (1+\beta\eta\,\cos\theta)\mpo{^2}} +\frac{1}{1-\eta^2} \ . 
\label{eq:xmodedisp}
\end{equation}
Here, $\eta^2=c^2 k^2 /\omega^2$ is the refraction index in the simulation frame and $\theta $ is the angle between the $x-$axis and the wave propagation direction.
\mpo{The first term on the right hand side is relevant only for $\eta\ll 1$, in which case it is approximately $\sigma/\gamma^2$. We can therefore always write it in that form and derive the simplified dispersion relation}
\begin{equation}
\omega \approx \sqrt{(1+\frac{\sigma_e}{\gamma^2})\omega_{pe}^2 + c^2 k^2} .
\label{eq:xmodedisp_approx_fin}
\end{equation} 
Precursor waves propagate toward the shock upstream. The cutoff wave number is then determined by equating {the $x$-component of the wave group velocity, $v_{\rm gr,x}= v_{\rm gr}\cos{\theta}$} with the shock velocity, $c\beta_{\rm sh}$, yielding:
\begin{equation}
k_x= \beta_{\rm sh}\gamma_{\rm sh}\sqrt{(1+\frac{\sigma_e}{\gamma^2})\frac{\omega_{pe}^2}{c^2} +k_y^2}. 
\label{eq:moded_cutoff}
\end{equation}

The cutoff wave number for O-mode \mpo{waves, that we use in Fig.~\ref{fig:fourier_B_ip},} can be estimated in an analogous way. The dispersion relation \mpo{in the simulation frame} is the same as for the electromagnetic wave in unmagnetized plasma ($\sigma=0$),
\begin{equation}
\omega^2 = k^2 c^2 + \omega_{pe}^2 \,
\label{eq:omodedisp}
\end{equation}
which leads to the cutoff wave number for the ordinary mode:
\begin{equation}
k_x= \beta_{\rm sh}\gamma_{\rm sh}\sqrt{\frac{\omega_{pe}^2}{c^2} +k_y^2} \ .
\label{eq:omoded_cutoff}
\end{equation}

\begin{figure}
\begin{center}
\includegraphics [width=1.0\linewidth] {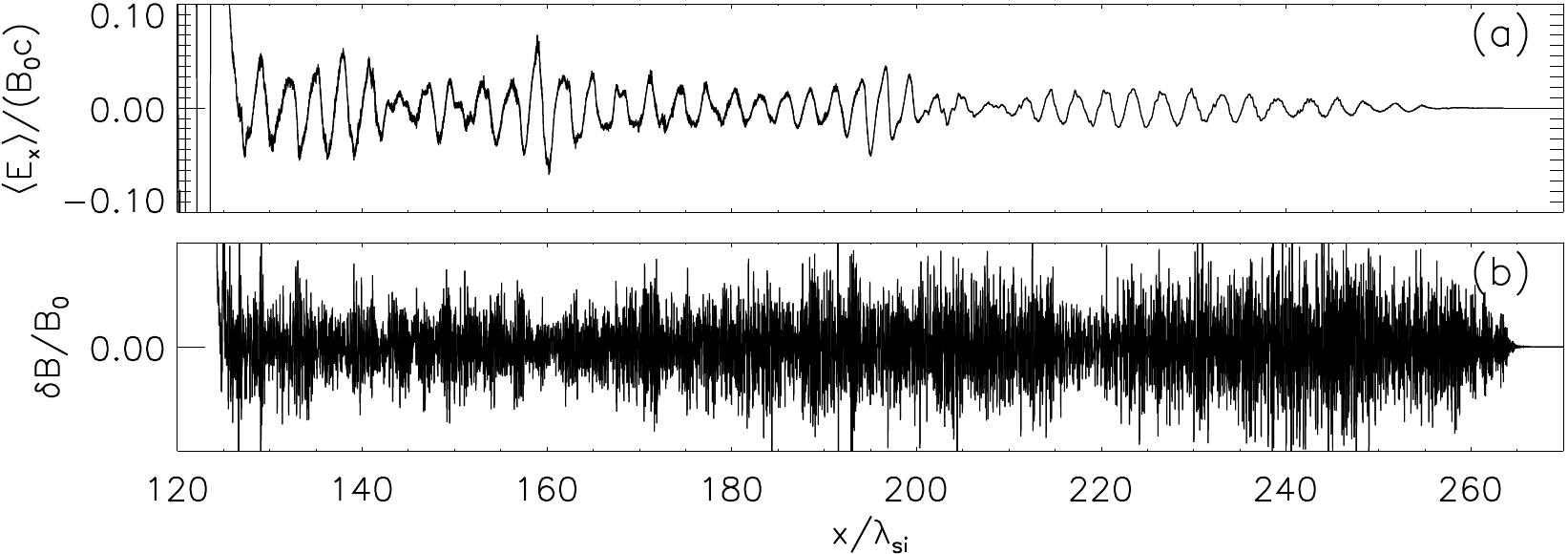}
\end{center}
\caption{Upstream wave profiles for $\langle\Ex\rangle$ (a) and $\delta\Bz$ (b) in a 1D simulation at time $t\Wci = 84.7$.}
\label{fig:zoom_1d}
\end{figure}

\section{Results for 1D simulation}
\label{app:1d}
\mpo{The 2D simulations presented in Sections~\ref{sec:out-of-plane} and~\ref{sec:in-plane} can be compared with a 1D run, in which shock rippling and obliquely emitted waves are} suppressed.  
In the ultra relativistic regime the amplitude of SMI-generated precursor waves are systematically larger \mpo{in 1D than in 2D, on account of the loss in phase coherence imposed by inhomogeneity in the shock surface \citep{iwamoto2017,iwamoto2018}.  
There is a positive feedback in electron-ion plasma, however, through} which electrons accelerated in the shock upstream enhance the precursor wave emission that in turn induce stronger wakefield, accelerating the incoming electrons even more efficiently, leading to energy equipartition between electrons and ions \citep{lyubarsky2006,hoshino2008}. \mpo{At ultrarelativistic shocks with high electron magnetizations, $\sigma_e \gtrsim 1$,
the amplitude of the precursor waves in 2D is comparable to that in 1D \citep{iwamoto2019}. In our 2D simulations of moderately relativistic shocks the positive feedback process is not operative, and the wave amplification may be attributed to shock rippling. A 1D test simulation is performed to evaluate these issues.}

\begin{figure}
\begin{center}
\includegraphics [width=1.0\linewidth] {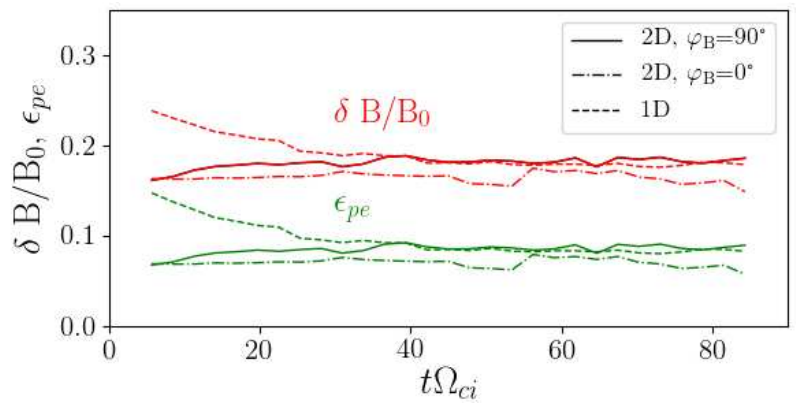}
\end{center}
\caption{Temporal evolution of normalized precursor wave amplitudes, $\delta B / B_0$ (red lines), and precursor wave energy normalized to the upstream electron kinetic energy, $\epsilon_{\rm pe}$ (green lines), for 2D out-of-plane (thick solid lines), 2D in-plane (dash-dotted-lines) and 1D simulations (dashed lines).}
\label{fig:ampl_a_oop}
\end{figure}

The setup of the 1D simulation is the same as that for 2D simulations, but the transverse dimension of the computational box is only 5 cells wide, \mpo{making it effectively 1D.} 
Fig.~\ref{fig:zoom_1d} {shows the wave} profile for the 1D {simulation at} time of $t\Wci = 84.7$, {which is $t_{max}$ in 2D simulations.} The shock \mpo{speed is similar to that} {measured in} the 2D out-of plane simulation.
\mpo{The fluctuations in $\Ey$ and $\Bz$ are anti-correlated, indicating that the waves are of X-mode type.}

\begin{figure}
\begin{center}
\includegraphics[width=1.0\linewidth]{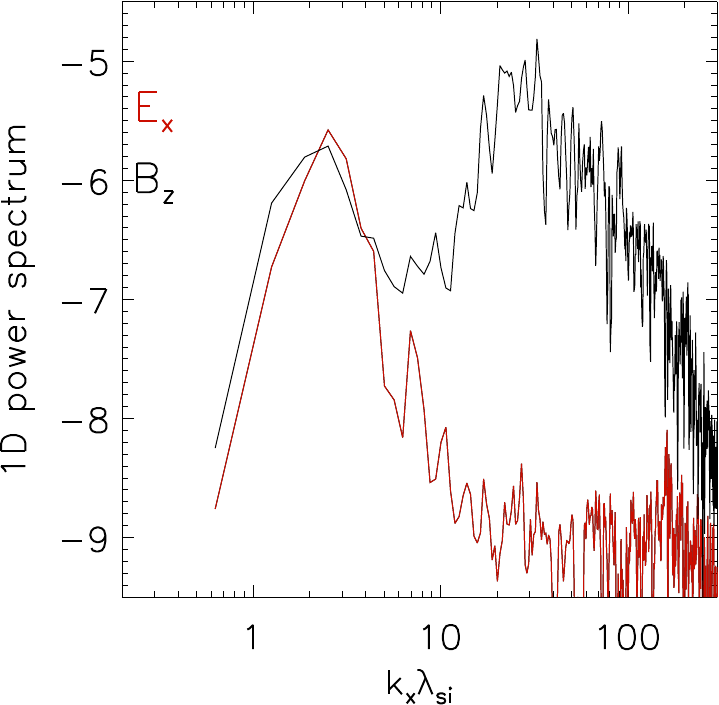}
\end{center}
\caption{1D power spectrum for $\Bz$ (black) and $\Ex$ (red) in the region $x/\Lsi = 130-140$ at time $t\Wci = 84.7$ (cf. Fig.~\ref{fig:zoom_1d}).}
\label{fig:four1d}
\end{figure}
The precursor wave {profiles}
in the 1D run can be directly compared with {those obtained in} 2D (see Fig.~\ref{fig:profile} and also Fig.~\ref{fig:shock_str_z_ip}). {The wave amplitude is also listed in Table~\ref{table1}.}
\mpo{Whereas in 2D wave amplification by the shock ripples 
causes a high wave amplitude near the shock, in the 1D case the {electrons heated upstream of the shock reduce} the later emission of the precursor waves whose amplitude is then low \citep{amato2006}. This behavior is also evident in the time evolution of the normalized wave amplitude, $\delta B/B_0$, and wave energy,
$\epsilon_{\rm pe}$, that we show in Fig.~\ref{fig:ampl_a_oop}.}

One can note that the wave evolution is similar in both 2D runs, and 
in the 2D in-plane case the total precursor waves amplitude is only slightly smaller than the one observed for $\varphi_B=90\degree$. This shows that the shock rippling-{mediated} wave amplification operates \mpo{with similar efficiency in both 2D setups, despite the} different mechanisms.

Fig.~\ref{fig:four1d} shows 1D Fourier power spectra upstream of the shock, in the region $x/\Lsi = (130-140)$. 
The signal band in the magnetic field {oscillations}, $\Bz$, is consistent with the SMI precursor waves observed in 2D simulations. The electrostatic component in $\Ex$ has a wavenumber of $k_{\Ex}\approx 2$, consistent with the theoretical wave number for SMI-generated wakefield.

% Don't change these lines
\bsp	% typesetting comment
\label{lastpage}
\end{document}